\documentclass[letterpaper,twocolumn,10pt]{article}
\usepackage{usenix}

\usepackage{tikz}
\usepackage{amsmath}

\usepackage{filecontents}

\usepackage[T1]{fontenc}
\usepackage{graphicx}

\usepackage{amsmath,amsfonts}
\usepackage{booktabs}
\usepackage{graphicx}

\usepackage{textcomp}
\usepackage{xcolor}

\def\BibTeX{{\rm B\kern-.05em{\sc i\kern-.025em b}\kern-.08em
    T\kern-.1667em\lower.7ex\hbox{E}\kern-.125emX}}

\usepackage{hyperref}       
\usepackage{url}            
\usepackage{xurl}
\usepackage[capitalise]{cleveref}
\usepackage{multirow}

\usepackage{enumitem}

\usepackage{algorithm}
\usepackage{algpseudocode}
\usepackage{xspace}

\algnewcommand{\Inputs}[1]{%
  \Statex \textbf{Inputs:}
  \Statex \hspace*{\algorithmicindent}\parbox[t]{1.\linewidth}{\raggedright #1}
}
\algnewcommand{\Outputs}[1]{%
  \Statex \textbf{Outputs:}
  \Statex \hspace*{\algorithmicindent}\parbox[t]{1.\linewidth}{\raggedright #1}
}
\algnewcommand{\Initialize}[1]{%
  \Statex $C_i$\ \textbf{initialize:}
  \Statex \hspace*{\algorithmicindent}\parbox[t]{1.\linewidth}{\raggedright #1}
}

\usepackage[
adversary,
asymptotics,
ff,
lambda,
landau,
keys,
mm,
operators,
primitives,
sets,
]{cryptocode}

\usepackage[capitalise]{cleveref}
\usepackage{multirow}

\usepackage{xspace}

\newcommand{\keygen}{\mathsf{KeyGen}}

\newcommand{\etal}{{\it et~al.}\xspace}
\newcommand{\ie}{{\it i.e.}\xspace}
\newcommand{\eg}{{\it e.g.}\xspace}
\newcommand{\cf}{{\it cf.}\xspace}
\newcommand{\resp}{resp.\xspace}

\newcommand{\usr}{\mathsf{U}\xspace}

\usepackage{tikz}
\usetikzlibrary{decorations.pathmorphing,arrows.meta,positioning}

\newcommand{\SIG}{\ensuremath{\mathsf{SIG}}\xspace}
\newcommand{\SSC}{\ensuremath{\mathsf{SSC}}\xspace}
\newcommand{\keystream}{\mathsf{Keystream}}
\newcommand{\AKE}{\ensuremath{\mathsf{AKE}}\xspace}

\usepackage{float}
\usepackage{subfig}

\newcommand{\RAMP}{\textit{RAMP}\xspace}
\usepackage{amsthm}
\theoremstyle{plain}
\newtheorem{definition}{\textbf{Definition}}
\newtheorem{theorem}{\textbf{Theorem}}
\newtheorem{lemma}{\textbf{Lemma}}
\newtheorem{corollary}{\textbf{Corollary}}

\mathchardef\mhyphen="2D
\newcommand{\indcpa}{\ensuremath{\mathsf{IND}\mhyphen\mathsf{CPA}}}

\usepackage{threeparttable}
\usepackage{pgfplots}
\pgfplotsset{compat=1.18}
\usepackage{flushend}

\begin{document}
\title{\Large \bf RAMP: Toward Ubiquitous Quantum-Safe Key Distribution as a Service}
\author{
{\rm Jiafan Wang}\\
CSIRO Technology
\and
{\rm Sebastian Kish}\\
University of Technology Sydney
\and
{\rm Dongxi Liu}\\
CSIRO Technology
} 

\maketitle              
\begin{abstract}

Quantum key distribution (QKD) offers a security foundation independent
of computational hardness assumptions. However, its reliance on specialized hardware limits access by ordinary users. 
Turning QKD into a service poses a fundamental trust challenge, as the providers that supply QKD-derived material themselves could retain knowledge of the keying inputs of users. This makes end-to-end confidentiality
solely depend on the trustworthiness of the service infrastructure.

We present \RAMP, a service-oriented framework that
decouples QKD material generation from key consumption. Providers
operate QKD infrastructure and supply raw-key materials,
while users securely acquire, mask, and locally process them into final
keys \emph{without operating QKD devices}. \RAMP employs
reconciliation-preserving masking compatible with both DV- and CV-QKD.
It mitigates the trust dependence by separating the QKD-service and user-side masking domains.

We formalize complementary confidentiality guarantees: \RAMP protects the final key even if colluding providers expose their complete raw-key views, provided the masking domain remains secure; conversely, exposing the complete masking state does not break final-key secrecy as long as the QKD source and acquisition paths remain protected. The results establish the dual-domain security boundary of \RAMP.

We implement \RAMP using experimentally acquired QKD data and standardized quantum-safe cryptography. After acquiring service-provisioned keying material, commodity endpoints complete user-side QKD post-processing with an average key-distillation throughput of $2.1$Kb/s on an M4 Pro CPU. 
To our knowledge, we demonstrate the first experimental realization of such service-provisioned, user-side QKD processing, presenting a viable path toward QKD as a service.

\end{abstract}

\section{Introduction}

Large-scale quantum computers would compromise widely deployed
public-key mechanisms such as RSA and Diffie--Hellman~\cite{focs/Shor94},
motivating the transition toward quantum-safe key establishment.
Post-quantum cryptography (PQC)~\cite{bernstein2017post} provides a software-based
migration path\footnote{\url{https://www.nccoe.nist.gov/applied-cryptography/migration-to-pqc}} for replacing public-key primitives vulnerable to quantum attacks, while quantum key distribution (QKD)~\cite{journals/tcs/BennettB14} offers a fundamentally distinct approach of secret generation 
based on quantum physics~\cite{wootters1982single}.
Appropriately parameterized
symmetric cryptography~\cite{chennistir} also remains a practical building block against
quantum adversaries once shared secret state has been established.

The deployment of PQC does not eliminate the value of QKD.
PQC remains computational, \ie, its security relies on the hardness of
mathematical problems against classical and quantum adversaries.
Instead, QKD derives its secrecy from the properties of
quantum communication under a specified physical model. The two paradigms thus provide \emph{complementary trust foundations for
quantum-safe key establishment}.

This distinction has motivated substantial investment in QKD
infrastructure and high-assurance communication systems\footnote{For example, secure quantum
communication backbone across the
European Union~(\url{https://digital-strategy.ec.europa.eu/en/policies/european-quantum-communication-infrastructure-euroqci}) and quantum information science as a strategic national priority in US (\url{https://www.quantum.gov/}).}.
However, its deployment model remains difficult to scale. 
QKD requires specialized optical hardware, calibrated quantum links, and protocol-specific
processing at dedicated nodes. Thus, it is more desirable to make this scarce infrastructure accessible to ordinary endpoints \emph{via a service-oriented model rather than simply deploying more QKD devices}.

\begin{figure*}[t]
\centering
\includegraphics[width=0.8\textwidth]{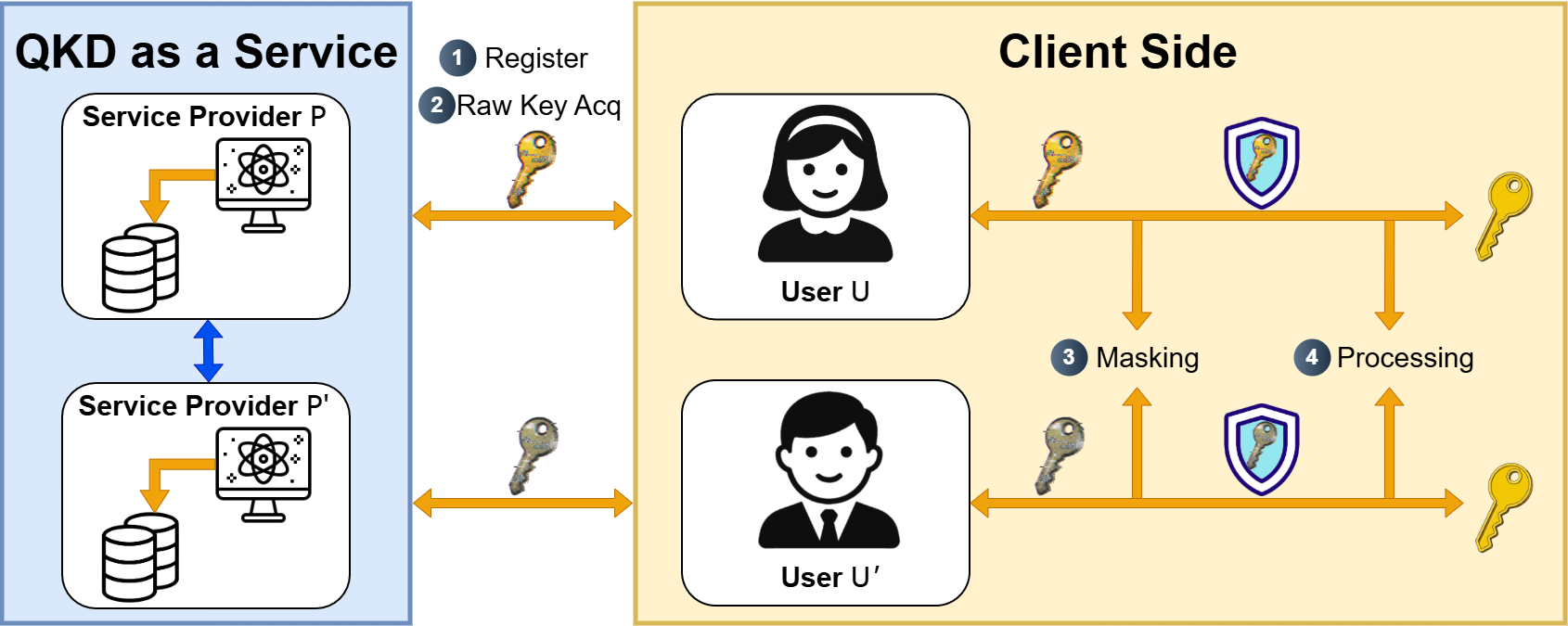}
\caption{The \RAMP Framework (Blue Arrow for Quantum Channel)} 
\label{system_model}
\end{figure*}

\subsection{Challenges in More Accessible QKD}
Existing QKD systems~\cite{journals/tcs/BennettB14,ralph1999continuous, inoue2002differential,xu2020secure} primarily operate in a point-to-point manner. 
The two-party key establishment requires dedicated optical links and a pair of QKD devices. 
Both parties participate in quantum-state transmission to generate \emph{raw keys}, and then perform classical post-processing (including sifting, parameter
estimation, information reconciliation, and privacy amplification) to distill the final secret key from raw keys.
It is physically and economically infeasible to deploy specialized hardware at numerous endpoints.

\smallskip\noindent 
\textit{Device-Centric QKD Networks.}
The proposals of QKD network~\cite{mehic2020quantum, journals/comsur/CaoZWZNH22} aim to scale point-to-point QKD by multiplexing quantum channels and coordinating multi-hop relaying for broader coverage.
National initiatives, such as SECOQC~\cite{peev2009secoqc}, SwissQuantum~\cite{stucki2011long}, and Tokyo QKD Network~\cite{sasaki2011field}, showcase the viability of linking multiple QKD nodes. 

While the advancement shows the potential to extend QKD networks beyond laboratory environments, these systems remain resource-intensive, imposing large overhead. 
Each node must host modules for quantum state transmission, maintain stable quantum channels, and handle post-processing. 
Note that the model of QKD network is inherently device-centric. 
It interconnects a pre-configured set of nodes, instead of offering on-demand key acquisition for regular users.

\smallskip\noindent 
\textit{Trust Issues in QKD as a Service.}
Recent works~\cite{cao2019multi, cao2019sdqaas, conf/ispec/KozlovicsPLV23} attempt to make QKD a service that is available to multiple users. 
Existing designs mainly focus on resource and service provisioning over deployed infrastructures of QKD networks, including secret-key assignment, network orchestration, and the integration of QKD-derived keys into classical protocols.

These efforts improve the operational flexibility of QKD networks, 
but their service boundary typically remains trusted:
the infrastructure that provisions a QKD-derived key may learn or
reconstruct the key delivered to the user. Consequently, compromise
or collusion within the service infrastructure can directly devastate
end-to-end key confidentiality.

We comprehensively review related work in~\cref{sec:related}.
The literature leaves open a fundamental question:
\begin{quote}
    \textit{Can QKD-derived keying material be delivered as a convenient service to users \textbf{without QKD hardware}, while \textbf{avoiding sole reliance} on the secrecy and trustworthiness of the service infrastructure for end-to-end final-key confidentiality?}    
\end{quote}

\subsection{From QKD to Quantum-Safe Service}
We answer the question affirmatively with a service-oriented framework called \RAMP. 
\RAMP 
enables users to obtain QKD-derived keys \emph{without owning specialized devices}, while \emph{limiting the trust placed in the service providers} that supply the underlying QKD materials. 
\RAMP denotes the user-side operational cycle, \ie,  \textit{Register–Acquire–Mask–Process}.

\RAMP users obtain and leverage the entropy from QKD by interacting with their service providers. 
We strategically separate the quantum and the classical stages of QKD: the service providers handle the transmission of quantum states and the maintenance of fresh raw-key materials, while the users perform the cryptographic protection and QKD post-processing
locally using general-purpose computing hardware.
This separation allows a shared QKD backend to supply raw-key
materials to multiple users without requiring QKD hardware at the
user endpoints, while decoupling continuous QKD generation from
on-demand key consumption.

\smallskip\noindent 
\textit{Technical Challenges and Solutions.}
The design of \RAMP addresses
several challenges in integrating QKD with standardized quantum-safe building blocks, while preserving correctness, effectiveness, and security in a service-oriented model.

\begin{itemize}
    \item \textbf{Complementary Compromise Resilience.}
    Moving QKD behind a service boundary creates an unusual security problem, as the providers by definition know the complete QKD material supplied to the users for secret derivation. \RAMP therefore introduces an independent user-side masking domain, such that provider compromise is tolerated through the masking domain, while exposure of the masking state falls back to the secrecy of the protected QKD source and acquisition paths.

    \item
    \textbf{Reconciliation-Preserving Masking.}
    As QKD raw-key materials are inherently noisy, conventional encryption cannot simply be inserted into the processing pipeline without considering its effect on reconciliation. \RAMP applies customized relabeling to raw keys via a masking mechanism. 
    This transformation preserves the underlying discrepancy or likelihood structure of general QKD modes, allowing standard QKD post-processing to operate directly on the masked representation.
    As a byproduct, \RAMP retains the underlying QKD soundness for detecting unilateral inconsistent or low-quality material.

    \item
    \textbf{Service-Oriented Key Acquisition.}
    The independently stored raw-key materials should be bound to the intended user session without requiring specific coordination between the service providers, in order to prevent unintended reuse.  \RAMP therefore 
    deploys user-mediated acquisition, enforcing identifier binding and raw-key lifecycle management before masking and post-processing.

    \item \textbf{Interoperability with Standardized Suites.}
    Compatibility with standardized implementations of QKD and PQC poses a substantial challenge to practical deployment. 
    \RAMP is designed for generic QKD protocols.
    The construction incorporates standard cryptographic interfaces, which can be instantiated with well-developed post-quantum and symmetric primitives.
\end{itemize}

\smallskip\noindent 
\textit{Distinction against Key Combiner.}
\RAMP is not designed as a replacement for generic hybrid key
combiners~\cite{conf/pqcrypto/DowlingHP20,
conf/pqcrypto/BrucknerRS23}, which combine independently protected secrets at final-key derivation.
These works, when applied to QKD-as-a-service, could let the service providers complete QKD post-processing, deliver a QKD key, and combine it at the users with independently established pre- or post-quantum keys as the final key.

\RAMP deliberately \emph{advances the service boundary} from final-key delivery to key processing.
After receiving the material, the users themselves perform parameter estimation, reconciliation, and privacy amplification. Our reconciliation-preserving
masking makes this user-side QKD processing possible
without allowing provider knowledge of the supplied material to
determine the final key. 
\RAMP therefore proposes a distinct \emph{trust--cost tradeoff} by removing security-critical processing from the provider-side trusted execution path.

\subsection{Contributions}

Our contributions span system design, service-oriented
scheme, dedicated security analysis, and empirical evaluation.

\begin{itemize}
    \item \textbf{QKD-Derived Key Distribution as a Service.}
    We propose \RAMP, a service-oriented framework that decouples QKD material generation from key consumption. Specialized providers operate the quantum infrastructure and continuously produce correlated raw-key materials, while ordinary users acquire and locally process these materials on demand without operating QKD devices. 

    \item \textbf{Dual-Domain Resilient Key Establishment.}
    We design \RAMP around two cryptographically separated security domains, namely, the QKD service domain and the user-side post-quantum masking domain. 
    The raw-key acquisition is protected by independently provisioned credentials, while a pair of users establishes the masking secret through authenticated post-quantum key establishment. 
    \RAMP thus prevents a sole assumption from protecting both the QKD-derived material and its mask.

    \item \textbf{Reconciliation-Preserving Post-Processing Pipeline.}
    By employing masking with no error propagation, \RAMP supports the standard post-processing of general QKD modes and preserves its inherent error tolerance. 
    \RAMP further utilizes this property to enable the detectability of inconsistent materials from any one of the providers.

    \item \textbf{Security under Complementary Compromise Modes.}
    We formalize two complementary security notions for \RAMP. We prove final-key confidentiality against colluding service providers that expose their complete raw-key views, assuming the user-side post-quantum masking domain remains secure; meanwhile, we show that exposing the masking secret does not invalidate final-key secrecy when the underlying QKD execution and acquisition path remains uncompromised. These results establish the dual-domain compromise resilience of \RAMP.

    \item \textbf{Evaluation with Real-World QKD.}
    We implement the \RAMP processing pipeline on realistic user devices, using experimentally acquired QKD data and standardized quantum-safe cryptographic primitives. 
    We demonstrate practical secret-key throughput with modest \RAMP-specific cost over the underlying QKD processing.
    
\end{itemize}

\smallskip\noindent 
\textit{Organization.}
\cref{sec:prelim} provides preliminaries on QKD and involved cryptographic primitives.
\cref{sec:system-design} presents our system model, 
including the interfaces, threat model, and security goals.
We construct \RAMP in~\cref{sec:construction} and formalize its security in~\cref{sec:security}.
\cref{sec:real_qkd_eval} reports our experimental evaluation of \RAMP and \cref{sec:conclusion} concludes this work.

\section{Preliminaries}
\label{sec:prelim}
\subsection{Quantum Key Distribution (QKD)}
QKD~\cite{journals/tcs/BennettB14} enables two distant parties, conventionally referred to as Alice and Bob, to establish a shared secret key using quantum communication. The secret is encoded into quantum states, and QKD security relies on fundamental properties of quantum mechanics, \eg, the no-cloning theorem~\cite{wootters1982single}. Any attempt by an adversary to intercept or measure the transmitted quantum states, regardless of its computational power, inevitably introduces disturbances, which can be detected by the legitimate parties. As a result, QKD provides information-theoretic security of the generated keys~\cite{PhysRevA.72.012332}.

We divide the QKD procedure into two stages: the \emph{quantum stage}, where quantum states are exchanged and measured, and the \emph{classical stage}, where post-processing operations are performed over an authenticated classical channel.

\smallskip\noindent 
\textbf{Quantum Stage.}
This stage consists of preparation, transmission, and measurement of quantum states over a quantum channel. Depending on how information is encoded and measured, QKD protocols are categorized as discrete-variable QKD (DV-QKD)~\cite{journals/tcs/BennettB14} and continuous-variable QKD (CV-QKD)~\cite{ralph1999continuous}. DV-QKD encodes with discrete quantum degrees of freedom (\eg, photon polarization), while CV-QKD uses continuous modulation of optical field quadratures.

At the end of the quantum stage, Alice and Bob obtain correlated classical data derived from their respective preparations and measurements. These data, referred to as \emph{raw keys}, are generally noisy and only partially correlated due to channel loss, noise, and potential eavesdropping. In addition to the raw keys, the devices record \emph{metadata} such as detection efficiency, detector electronic noise, shot-noise calibration, timing statistics, and other implementation parameters used for subsequent parameter estimation and security evaluation.

\smallskip\noindent 
\textbf{Classical Stage.}
After quantum transmission, Alice and Bob perform classical post-processing over an authenticated channel to distill a shared secret key. Although specific implementations differ between DV-QKD and CV-QKD, the classical stage generally consists of the following steps:

\begin{itemize}
\item 
Data Filtering (Sifting):
Alice and Bob publicly exchange protocol-dependent information (\eg, basis choices in DV-QKD or measurement/configuration information in CV-QKD) to identify and retain subsets of data that are sufficiently correlated for key generation. 
This step may be implicit or combined with other processing.

\item
Parameter Estimation:
A subset of the data is disclosed to estimate channel characteristics and noise levels. DV-QKD typically quantifies it by the quantum bit error rate (QBER), while CV-QKD involves estimating quantities such as transmission, excess noise, and signal-to-noise ratio (SNR). If the estimated parameters exceed security thresholds, the protocol is aborted.

\item
Information Reconciliation:
Due to noise and imperfections, Alice and Bob’s raw keys are not identical. Error correction techniques are applied to align their data. Interactive protocols such as Cascade~\cite{conf/eurocrypt/BrassardS93} are commonly used in DV-QKD, while forward error correction using low-density parity-check (LDPC) codes~\cite{LDPC} is widely used in CV-QKD and high-rate systems.

\item
Privacy Amplification:
To remove any partial information potentially available to an adversary, Alice and Bob apply universal hash functions to compress the reconciled key into a shorter, uniformly random secret key. The final key length depends on the estimated leakage during reconciliation and the security parameters.

\end{itemize}

Alice and Bob end with an identical shared secret key that is provably secure under the chosen QKD protocol.

\newcommand{\Aux}{\ensuremath{\mathsf{Aux}}\xspace}

\smallskip\noindent 
\textit{Composable QKD Secrecy.}
We adopt the standard composable secrecy notion of QKD~\cite{journals/corr/PortmannR14,journals/quantum/TomamichelL17}.
Let $K\in\{0,1\}^{\ell}$ denote the final key produced by an accepted QKD execution, and let $E$ denote the adversary's complete side-information system allowed by the instantiated QKD security model. 
In particular, $E$ may contain quantum information retained from the quantum-channel interaction together with the public classical post-processing transcript.

Let $\bar{\rho}^{\mathsf{acc}}_{KE}$ denote the normalized
classical--quantum state of $K$ and $E$ conditioned on protocol
acceptance. 
Let
$\bar{\rho}^{\mathsf{acc}}_{E}
 = \operatorname{tr}_{K}
   \bar{\rho}^{\mathsf{acc}}_{KE}$, where $\operatorname{tr}_{K}$ denotes the partial trace over $K$, \ie, $\bar{\rho}^{\mathsf{acc}}_{E}$ is the marginal state of $E$ obtained by discarding $K$.
The ideal key state is
$$
\tau_K
=
2^{-\ell}
\sum_{k\in\{0,1\}^{\ell}}
|k\rangle\langle k|,
$$
which represents a uniformly distributed $\ell$-bit
key. For quantum states $\rho$ and $\sigma$, we write
$$
D(\rho,\sigma)
=
\frac{1}{2}\|\rho-\sigma\|_1
$$
for their trace distance. When the systems are entirely classical,
$D$ reduces to the usual total-variation (statistical) distance.

Let $p_{\mathsf{acc}}$ denote the probability that the QKD
execution accepts. The protocol provides
$\varepsilon_{\mathsf{qkd}}$-secrecy if, for every attack covered by
the instantiated QKD security model,
$$
p_{\mathsf{acc}}\,
D\!\left(
\bar{\rho}^{\mathsf{acc}}_{KE},
\tau_K\otimes\bar{\rho}^{\mathsf{acc}}_{E}
\right)
\le
\varepsilon_{\mathsf{qkd}}.
$$
Operationally, this bounds the distinguishability between the real
QKD key and an ideal uniform key that is independent of all
adversarial side information.

\subsection{Cryptographic Building Blocks}
\label{sec:qfc}
We review the syntax of cryptographic building blocks used in \RAMP.
\cref{appx:pqc} recalls their security notions and discusses their quantum resistance, respectively.

\subsubsection{Digital Signature Scheme (SIG)}
A digital signature scheme~\cite{books/crc/KatzLindell2014} enables a party to 
authenticate messages and prove their origin. 
A signer generates a signature on a message using a secret signing key, and anyone could use the corresponding public key to verify the authenticity and integrity of the message. 

\begin{definition}[Digital Signature]
\label{def:SIG}
A digital signature scheme $\SIG = (\keygen,\allowbreak\sign,\verify)$ 
with message space $\mathcal{M}$ and signature space $\mathcal{S}$ 
is defined as follows:
\begin{itemize}
    \item $(pk,sk)\gets\keygen(\secparam)$: The algorithm takes as input a security parameter $\secparam$. 
    It outputs a secret signing key $sk$ and its corresponding public key $pk$. 
    \item $\sigma \gets \sign(sk,m)$: The signing algorithm outputs a signature $\sigma \in \mathcal{S}$ with a secret key $sk$ and a message $m\in\mathcal{M}$. 
    \item $b \gets \verify(pk,m,\sigma)$: The verification algorithm takes as input a public key $pk$, a message $m\in\mathcal{M}$, and a signature $\sigma\in\mathcal{S}$. 
    It outputs $b \in \{0,1\}$, where $b=1$ indicates $\sigma$ is a valid signature on $m$; $b=0$, otherwise. 
\end{itemize}
\end{definition}

\subsubsection{Authenticated Key Exchange (AKE)}
An authenticated key-exchange protocol~\cite{conf/crypto/BellareR93,conf/eurocrypt/CanettiK01} allows two parties to establish a shared session key while authenticating the intended communicating peer.

\begin{definition}[Authenticated Key Exchange]
An authenticated key-exchange protocol
\AKE is a two-party, session-based protocol with key space $\mathcal{K}$ executed
over an adversarially controlled network.
A local session at party $U$ with intended peer $U'$ is identified by
$\pi=(\usr,\usr',\mathsf{sid},\mathsf{role})$,
where $\mathsf{role}\in\{\mathsf{init},\mathsf{resp}\}$ and $\mathsf{sid}$ is a session identifier.

Upon successful completion on two matching sessions $\pi=(\usr,\usr',\mathsf{sid},\mathsf{init})$ and
$\pi'=(\usr',\usr,\mathsf{sid},\mathsf{resp})$, $\AKE(\usr,\usr',\mathsf{sid})$ outputs a session key
$k\in\mathcal{K}$; otherwise, it outputs $\bot$.

\end{definition}

\subsubsection{Secure Channel from Provisioned Credential}

A secure channel (SC) protects communication between two parties over an
adversarially controlled network by providing confidentiality and
integrity of the transmitted messages~\cite{conf/crypto/Krawczyk01}.
A secure channel from provisioned credentials lets the communicating parties establish a long-term symmetric credential via a prior authenticated enrollment or provisioning phase.

The provisioned credential serves as the trust root for subsequent
protected sessions. In particular, the online channel protection does
not require an additional public-key trust assumption.
The provisioned credential is subsequently used to authenticate
the peers and derive session-specific traffic keys for protected
communication.
This model is supported by standardized protocols. In particular,
TLS~1.3 enables such a channel based on symmetric pre-shared keys
(PSKs)~\cite{rfc9257,rfc9846}.

\subsubsection{Synchronous stream ciphers (SSC)}

Synchronous stream ciphers (SSC)~\cite{books/crc/MenezesOV96} generate the keystream independently of the plaintext and ciphertext, using only a secret key (and possibly an initialization vector). 
Encryption (\resp Decryption) simply executes bit-wise XOR between the plaintext (\resp ciphertext) and the keystream.

It implies that 
any transmission errors in the ciphertext (\resp plaintext) \emph{affect only the corresponding symbols} in the plaintext (\resp ciphertext) during operations. 
This property is also referred to as \emph{no error propagation}.   

\begin{definition}[XOR-Based Synchronous Stream Cipher]
\label{def:SSC}
An XOR-based synchronous stream cipher $\SSC=(\keygen,\keystream,\enc,\dec)$ with key space $\mathcal{K}$
is defined as follows:
\begin{itemize}
    \item $k \gets \keygen(\secparam)$: The key generation algorithm
    inputs a security parameter $\secparam$ and outputs a secret key $k \in \mathcal{K}$.
    \item $z \gets \keystream(k,n)$: The keystream generation algorithm takes as input a secret key $k$ and an output length $n$. 
    It outputs a $n$-bit keystream $z\in\{0,1\}^n$. 
    \item $c \gets \enc(m, z)$: 
    The encryption algorithm takes as input a message $m\in\{0,1\}^n$ and a keystream $z\in\{0,1\}^n$.
    It outputs the ciphertext $c= m \oplus z$, where $\oplus$ denotes the bit-wise XOR operation.
    \item $m \gets \dec(c, z)$: 
    The decryption algorithm takes as input a ciphertext $c\in\{0,1\}^n$ and the keystream $z\in\{0,1\}^n$. 
    It outputs the plaintext $m= c \oplus z$.
\end{itemize}
\end{definition}

\section{System Design}
\label{sec:system-design}
We aim to build a service-oriented system that enables end
users or applications to establish secret keys from QKD-derived entropy without deploying or operating specialized QKD hardware locally.
Instead, users interact with their service providers that maintain the QKD infrastructure and continuously generate raw-key materials.

The design follows a four-step workflow for any user:
\begin{itemize}
    \item 
    \emph{Register} with a local service provider for the key distribution service and establish the necessary credentials;
    \item 
    \emph{Acquire} raw keys from the service provider, which are continuously generated through QKD devices;
    \item
    \emph{Mask} the acquired raw keys to protect them from provider-side exposure and enhance security; and
    \item 
    \emph{Process} the masked raw keys through post-processing to derive the final distilled secret key.
\end{itemize}

We assume that the service providers manage their own physical QKD devices. 
These devices feature the basic functionality of regular point-to-point QKD protocols (either DV- or CV-QKD) when being initialized and paired.

As \cref{system_model} depicts, our description focuses on the scenario where two users subscribe to the service from their local providers.
The interfaces naturally admit deployments with multiple providers and subscribers, which is not evaluated in this work due to the resource limit.

\subsection{Interfaces}
\label{subsec:interface}

\newcommand{\init}{\mathsf{Init}\xspace}
\newcommand{\Regist}{\mathsf{Regist}\xspace}
\newcommand{\RKGen}{\mathsf{RKGen}\xspace}
\newcommand{\RKAcq}{\mathsf{Acq}\xspace}
\newcommand{\RKProc}{\mathsf{Proc}\xspace}
\newcommand{\Mask}{\mathsf{Mask}\xspace}
\newcommand{\spset}{\mathcal{P}\xspace}
\newcommand{\usrset}{\mathcal{U}\xspace}
\renewcommand{\sp}{\mathsf{P}\xspace}
\newcommand{\kid}{{kid}\xspace}
\newcommand{\rk}{{rk}\xspace}
\newcommand{\md}{{md}\xspace}
\newcommand{\rkm}{\rk_\mathsf{mask}\xspace}
\newcommand{\kfinal}{k_\mathsf{final}\xspace}

\newcommand{\kidset}{\mathcal{KID}\xspace}
\newcommand{\rkdict}{\mathcal{RK}\xspace}
\newcommand{\metadata}{\mathcal{MD}\xspace}

\newcommand{\kacq}{{k_{\usr,\sp}}\xspace}

\newcommand{\sid}{{sid}\xspace}

\noindent
\textbf{Notations.}
We denote the set of all service providers and the set of potential users/subscribers by $\spset$ and $\usrset$, respectively.
$\rkdict$ is a collection of raw-key materials and $\metadata$ records the corresponding metadata.
Both $\rkdict$ and $\metadata$ are indexed by a set of key identifiers $\kidset$.
For example, we can obtain a raw key $\rk$ from its collection via 
$\rk\gets\rkdict[\kid]$ with its key identifier $\kid\in\kidset$.

Without loss of generality, we use $\sp,\sp'\in\spset$ to denote two service providers that serve two users $\usr,\usr'\in\usrset$, respectively. 
We denote the holder of some data by its subscript,
\eg, $\kidset_\sp$ is the key identifiers by $\sp$, and omit it when it is clear. 

We define the interfaces of our system as below, each of which invokes a protocol by specified parties.

\begin{itemize}
    \item $\init(\spset)$: 
    All service providers from $\spset$ jointly initialize the system. 
    They build connections among pairs of QKD devices and confirm the specification.
    Each party sets up and announces public parameters (to potential users), \eg, information on certificates,  public keys, available service, and the data format.
    \item $\RKGen(\sp, \sp')$: 
    It invokes continuous QKD sessions between service providers 
    $\sp, \sp'\in \spset$ for raw-key generation.  
    These materials are stored as $\rkdict_\sp$ and $\rkdict_{\sp'}$, indexed by the corresponding key identifiers, and attached with the necessary metadata $\metadata_\sp$ and $\metadata_{\sp'}$, by both parties, respectively.
    \item $\Regist(\usr, \sp)$: User $\usr$ subscribes to the service from the local provider $\sp$.
    It provisions an independent symmetric acquisition credential $\kacq$, which is used exclusively to protect the raw-key acquisition between $\usr$ and $\sp$. 
    \item $\RKAcq(\usr, \sp, \kacq, \kidset_\sp, \rkdict_\sp, \metadata_\sp)$: 
    User $\usr$ acquires raw keys from the service provider $\sp$ where $\usr$ has registered. 
    The returned raw key is denoted as $\rk$.
    \item $\Mask(\usr, \usr', \rk, \rk')$: Two users $\usr, \usr'\in\usrset$ input $\rk$ and $\rk'$, respectively.
    $\rk$ and $\rk'$ are obtained via $\RKAcq()$ from $\sp$ and $\sp'$ with the same $\kid$, but may be different from each other due to channel noise or misbehavior.
    $\usr$ and $\usr'$ mask $\rk$ and $\rk'$, respectively, using user-derived keystreams.
    The masked raw keys are denoted as $\rkm$ and $\rkm'$, respectively.
    
    \item $\RKProc(\usr, \usr', \rkm, \rkm')$:
    Two users $\usr, \usr'\in\usrset$ conduct post-processing on $\rkm$ and $\rkm'$, yielding a final secret key $\kfinal$ shared only between them.
\end{itemize}

\subsection{Threat Model}
\label{subsec:threat-model}

\smallskip\noindent 
\textit{Adversary Capabilities.}
We consider a probabilistic or quantum polynomial-time adversary \adv\ that may control the classical network and exercise the capabilities permitted to an eavesdropper by the security model of QKD. In particular, \adv\ may intercept, delay, replay, drop, or modify classical messages exchanged among service providers and users, while interacting with the quantum channel as considered in QKD.

We expand the threat model to cover two forms of compromise that affect distinct parts of the QKD service system.
\begin{itemize}
\item 
\textbf{Service-provider corruption.} 
\adv\ may corrupt one or both service providers. Upon corruption, it obtains the provider's state associated with the target session, including raw-key materials, metadata, and acquisition credentials. 
\adv\ may deviate the provider from the prescribed protocol. This captures both curious providers that expose their local views and actively misbehaving providers that return manipulated or low-quality materials.
\item 
\textbf{Masking-state exposure.} 
\adv\ may obtain the user-side masking state of the target session, including the cryptographic session secret, the derived masking key, and the resulting masking keystream. This models compromise of the masking mechanism, such as exposure of its session secret or failure of its underlying public-key assumptions, either post-quantum or classical ones. 
\end{itemize}

 Users are considered honest, 
 as our target is to establish secret keys for them; the keys are otherwise trivially exposed.

\smallskip\noindent 
\textit{Security Conditions.}
We determine three security conditions that are associated with a target session of our system.
\begin{itemize}
  \item \textbf{Masking security (M).} 
  The secret state used to generate the user-side masking key remains uncompromised. Concretely, \adv\ does not obtain the masking secret, any derived masking keys, or the resulting keystream, and the involved cryptographic mechanisms remain secure.

  \item \textbf{QKD-service secrecy (Q).} 
  The underlying QKD execution satisfies its assumptions of secrecy and authenticity, and the provider-side state containing the target QKD raw-key materials is not exposed to \adv. 
  \textbf{Q} captures the authenticity of the user-handled QKD-related information and the secrecy of the QKD-derived source.
  
  \item \textbf{Acquisition security (A).} The acquisition paths for both users preserve the confidentiality and integrity 
  of the raw-key materials and their metadata, and the corresponding acquisition credentials remain uncompromised.
\end{itemize}

The conditions $\mathbf M$, $\mathbf Q$, and $\mathbf A$ do not
assume that compromise events are statistically independent in a
real deployment. They intend to identify distinct cryptographic assets and
trust roots whose exposure has different consequences.
Specifically, the acquisition credentials can be provisioned independently for each user--provider relationship and are not derived from the
user-to-user state. The user-side masking keystream is not derived from provider-side QKD or acquisition state. 
A complete compromise of user endpoints may expose several of
these assets simultaneously, which is outside the complementary-compromise guarantees in this work.

\smallskip\noindent 
\textit{Security goals.}
We target final-key confidentiality whenever either the masking condition remains secure or both the QKD-service and acquisition conditions remain secure, namely,
$$\mathsf{Conf}=\mathbf{M}\vee(\mathbf{Q}\wedge\mathbf{A}).$$
Upon protocol acceptance, we capture dual-domain resilience and inconsistency detectability with the following properties.

\begin{itemize}
  \item \textbf{Provider-Compromise Final-Key Confidentiality.} 
  Even if both service providers are corrupted and \adv\ obtains their complete views of the target session,
  the final key remains computationally indistinguishable from uniform, provided the user-side masking condition \textbf{M} holds.
  
  \item \textbf{Masking-Compromise Final-Key Confidentiality.} 
  Even if \adv\ obtains the complete masking state of the target session, the final key remains secret, provided that the QKD-service secrecy condition \textbf{Q} and the independent acquisition security condition \textbf{A} both hold.

  \item \textbf{Detectability of Inconsistent Key Material.}
  If one service provider supplies material that is inconsistent or otherwise inadmissible under the QKD post-processing procedure, \RAMP preserves the underlying QKD bound on the probability that such material would be accepted.

\end{itemize}

\cref{sec:security} formalizes these properties for our construction.

We assume 
that paired QKD devices could produce correlated raw keys with a statistical error rate that is observable via parameter estimation. 
The authenticity regarding (the public keys of) any party can be established via standard PKI of the TLS protocol~\cite{pqcrypto/BindelHMS17,ndss/SikeridisKD20}.
These assumptions concern operational requirements, orthogonal to the new challenges of our system.

\section{Construction}
\label{sec:construction}
\RAMP follows the four-step workflow, \ie, \emph{Register--Acquire--Mask--Process}.
As per the interfaces defined in~\cref{subsec:interface},
we separate 
system initialization $\init()$, 
continuous raw-key generation by providers $\RKGen()$, user registration $\Regist()$,
raw-key acquisition from providers $\RKAcq()$, 
keystream-based masking $\Mask()$, 
and post-processing between users $\RKProc()$.

\RAMP utilizes the following cryptographic functionalities. A digital signature scheme \SIG authenticates provider-originated service information. A secure acquisition channel between each user and its local provider is established via a provisioned symmetric credential.
Users establish a session secret through an authenticated key exchange protocol \AKE.  
The synchronous stream cipher \SSC expands this secret into a masking keystream.
The acquisition credential and \AKE secret are generated independently.
The key sizes and/or output lengths of involved primitives follow the suggestions of NIST~\cite{chennistir} for security against quantum adversaries.

\subsection{System Setup and Raw-Key Generation}
\label{cons:setup}
\smallskip\noindent 
\textit{$\init().$} 
All service providers from $\spset$ jointly bootstrap the global environment by registering service-provider identities, agreeing on parameters, setting up cryptographic suites, and establishing connections (provisioning physical QKD links).

\newcommand{\ST}{\mathcal{ST}\xspace}

\begin{enumerate}[leftmargin=2em]

  \item \textbf{Storage Initialization.} 
  Each service provider initializes its local storage, including the collection of raw keys $\rkdict$, metadata $\metadata$, and their index set $\kidset$. 
  It also maintains a collection $\ST$ to record the state of each raw-key block. 
  Depending on the usage of the material regarding $\kid\in\kidset$, 
  $\ST[\kid]\in$ 
  \[\{\text{AVAILABLE, PENDING, CONSUMED, EXPIRED}\}.\]
  The state of a pending raw-key block will eventually become \text{CONSUMED} or \text{EXPIRED}. 
  Once a raw-key block has been consumed or conservatively expired, 
  \RAMP never makes it available to any user session.
  
  \item \textbf{Digital Signature Setup.} 
  Each service provider executes $\SIG.\keygen(\secparam)$ to obtain its own private signing key and public verification key of \SIG.

  \item \textbf{QKD Connection.} Each service provider publishes a QKD capabilities descriptor, including the supported QKD mode (DV/CV), the structure of raw-key materials, the maximum tolerated error, and the metadata format.

  \item \textbf{Announcement.}
  The system publishes identities and public keys of all service providers, along with the routing data, as a reference for subscription. 
\end{enumerate}

\smallskip\noindent 
\textit{$\RKGen()$.}
Assume that 
$\sp, \sp'\in\spset$ operate a paired QKD link.
\begin{enumerate}
  \item \textbf{QKD Execution (Quantum Stage).} 
  $\sp$ and $\sp'$ continuously execute the quantum stage of the selected QKD protocol to obtain correlated raw keys. 
  Both providers assign each pair of correlated raw-key blocks a common identifier $\kid$ by hashing their own identities and a unique block identifier from the underlying QKD execution.
  \item \textbf{Storage Update.} 
  $\sp$ denotes the raw-key material and the metadata regarding $\kid$ by $\rk_\sp$ and $\md_\sp$, respectively.
  $\sp$ updates $\rkdict_\sp[\kid]$ as $rk_\sp$, $\metadata_\sp[\kid]$ as $\md_\sp$, and $\ST_\sp[\kid]$ as \textsf{AVAILABLE}.
  $\sp'$ updates on $\rk_{\sp'}$ and $\md_{\sp'}$.
\end{enumerate}

\begin{figure}[t]
\centering
    \centering
    \includegraphics[width=0.47\textwidth]{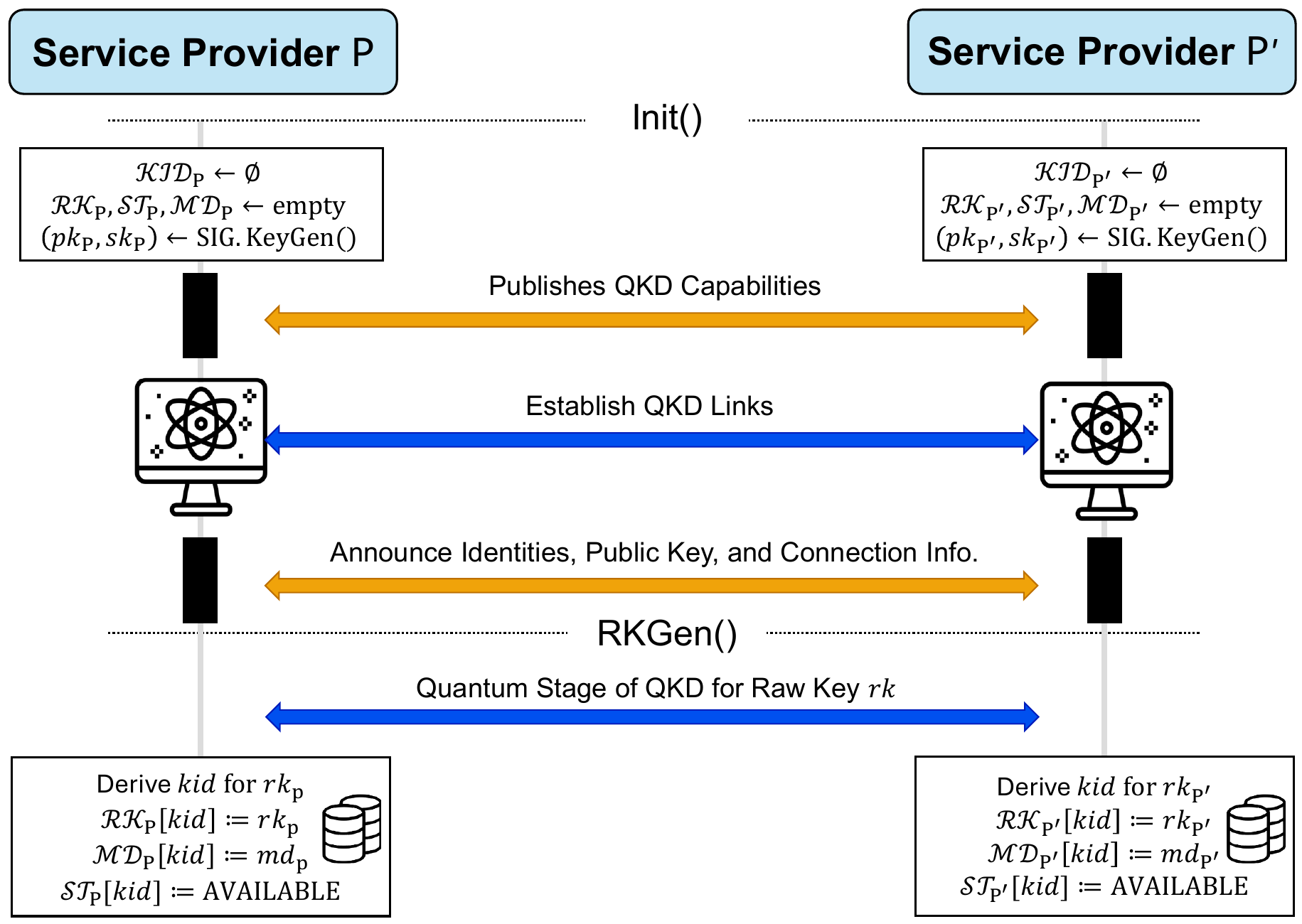}
\caption{System Setup and Raw-Key Generation}
\label{fig:setup}
\end{figure}

\cref{fig:setup} shows the system initialization and raw-key generation with $\spset=\{\sp, \sp'\}$.
The service providers could continuously generate and store raw-key material via $\RKGen()$.

\subsection{Registration and Raw-Key Acquisition}
\label{cons:register}

\smallskip\noindent 
\textit{$\Regist()$.} Enroll user $\usr\in\usrset$ with a local service provider $\sp\in\spset$ and establish a credential used exclusively by them.

More specifically, $\usr$ and $\sp$ provision an independent symmetric acquisition credential
$k_{\mathsf{acq}}^{\usr,\sp}\sample\{0,1\}^{\lambda}$
through an authenticated enrollment mechanism, which could be conducted out-of-band via provisioned or device-bound credentials~\cite{journals/rfc/rfc6063}.
$\usr$ and $\sp$ utilize $k_{\mathsf{acq}}^{\usr,\sp}$ to instantiate a secure communication channel 
for subsequent raw-key acquisition between them.

\smallskip\noindent 
\textit{$\RKAcq()$.} 
Consider that $\usr$,
served by $\sp$, 
intends to obtain a QKD-derived key with $\usr'$, 
served by $\sp'$.
The high-level idea of 
$\RKAcq()$ is to let $\usr$ acquire $\rk_\sp$ from local $\sp$ and forward $\kid$ of $\rk_\sp$ to $\usr'$, such that $\rk_{\sp'}$ regarding $\kid$ from $\sp'$ can be retrieved.
The following design aligns with standard QKD interfaces formalized by ETSI~\cite{ETSIAPI}.
After raw-key acquisition, 
$\usr$ and $\usr'$ hold correlated raw-key materials $\rk_\sp$ and $\rk_{\sp'}$ indexed by the same $\kid$, respectively. 
\begin{enumerate}
  \item $\usr \rightarrow \sp$: 
  $\usr$ requests raw-key material from $\sp$ through the acquisition channel between them and identifies $(\usr', \sp')$ as the intended peer user and its service provider.
  \item When receiving a valid request from a registered user, $\sp$ picks a key identifier $\kid\gets\kidset$ such that $\ST_\sp[\kid]=$AVAILABLE.
  $\sp$ retrieves the corresponding raw key 
  and its associated metadata as
  $$\rk_\sp:=\rkdict_\sp[\kid], \quad \md_\sp:=\metadata_\sp[\kid].$$
  $\sp$ updates $\ST_\sp[\kid]$ as \textsf{PENDING} to avoid race conditions.
  To bind the selected identifier to the intended user and provider pair, $\sp$ produces a signature as 
  $$
    \sigma_{\sp,\kid}
    \gets
    \SIG.\sign
    \left(
    sk_\sp,
    \texttt{Acq}
    \parallel \usr
    \parallel \usr'
    \parallel \sp
    \parallel \sp'
    \parallel \kid
    \right).
   $$

  \item $\sp \rightarrow \usr$: $\sp$ returns $(\rk_\sp,\md_\sp,\kid,\sigma_{\sp, \kid})$ to $\usr$.
  \item $\usr \rightarrow \usr'$: $U$ stores $(\rk_\sp,\md_\sp)$ and
  sends $\usr'$ $(\kid,\sigma_{\sp, \kid})$.
  \item $\usr' \rightarrow \sp'$: $\usr'$ submits $(\kid,\sigma_{\sp, \kid})$ to $\sp'$.
  \item
  Upon receiving $(\kid,\sigma_{\sp, \kid})$, 
  $\sp'$ verifies $\sigma_{\sp, \kid}$ with the public key $pk_\sp$ of $\sp$, and then checks whether $\ST_{\sp'}[\kid]=$AVAILABLE.
  If either check fails, the acquisition aborts.
  Otherwise,
  $\sp'$ updates $\ST_{\sp'}[\kid]$ as \textsf{PENDING} and retrieves from its local storage 
  $$\rk_{\sp'}:=\rkdict_{\sp'}[\kid],\quad \md_{\sp'}:=\metadata_{\sp'}[\kid].$$
  \item $\sp' \rightarrow \usr'$: 
  $\sp'$ returns $(\rk_{\sp'},\md_{\sp'})$ to $\usr'$.
\end{enumerate}

Before masking, $\usr$ and $\usr'$ respectively validate $(\rk_\sp,\md_\sp)$ and $(\rk_{\sp'},\md_{\sp'})$ against the QKD profile agreed during
initialization. Invalid or unsupported metadata causes abort.

\subsection{Raw-Key Masking}
\label{sec:mask}
\smallskip\noindent 
\textit{$\Mask()$.} 
$\usr$ and $\usr'$ now hold the raw-key materials $(rk_\sp,md_\sp)$ and $(rk_{\sp'},md_{\sp'})$, respectively, associated with the same identifier $\kid$. They then derive a common secret mask and apply it to the protocol-specific reconciliation inputs. No communication with $\sp$ and $\sp'$ would be required during this phase.

\begin{enumerate}
  \item \textbf{Session Binding.} 
  $\usr$ and $\usr'$ determine the session identifier $\sid$ by hashing $(U, U', P, P', \kid)$. 
  $\sid$ is used throughout the remainder of the user-side execution, binding the $\AKE$ session and subsequent QKD processing.

  \item \textbf{Pre-Mask QKD Preparation.} 
  The representation delivered by the QKD devices is protocol-dependent. Before masking, $\usr$ and $\usr'$ perform the public QKD operations that must precede reconciliation, including the applicable sifting and parameter-estimation steps. 
  $\usr$ and $\usr'$ authenticate the classical messages required by QKD. Any samples disclosed for parameter estimation are removed.
  
  \textit{Data Filtering (Sifting).} 
  $\usr$ and $\usr'$ obtain metadata (\eg, basis information in DV-QKD or measurement/configuration data in CV-QKD) 
  from $\md_\sp$ and $\md_{\sp'}$, respectively. 
  $\usr$ and $\usr'$ retain only the subset of data that is expected to be correlated. 

  \textit{Parameter Estimation.} 
  $\usr$ and $\usr'$ disclose publicly a random subset of filtered $\rk_\sp$ and $\rk_{\sp'}$ to estimate channel characteristics (\eg, QBER in DV-QKD or SNR in CV-QKD). 
  The step aborts if the result exceeds protocol-defined thresholds, indicating excessive noise or interference. 
  Since parameter estimation is performed only after both raw-key blocks have been acquired by the users, a unilateral provider cannot subsequently adapt its supplied block to the other user's disclosed samples.

  We denote the remaining portions on $\usr$ (\resp $\usr'$) after the above two operations by $x$ (\resp $x'$). $(x, x')$ is the input for subsequent reconciliation based on the original acquired $(\rk_\sp, \rk_{\sp'})$. 
  For DV-QKD, $x$ and $x'$ are binary strings retained for reconciliation. For CV-QKD, the preparation includes the protocol-specific conversion needed to obtain the binary reference string and the associated soft information used by reconciliation.
  
  \item \textbf{Masking-Key Establishment.} 
    After the reconciliation inputs $(x, x')$ have been fixed, $\usr$ and $\usr'$ establish a session-specific masking key through
    $$
    k_\AKE
    \gets
    \AKE(\usr,\usr',\sid).
    $$
    Masking proceeds only if the $\AKE$ execution accepts for the intended peer and the target $\sid$~\cite{conf/eurocrypt/CanettiK01}.

  \item \textbf{Keystream Derivation.}
    Let $n$ denote the length of the reconciliation input. 
    $\usr$ and $\usr'$ respectively derive an $n$-bit masking keystream bound to the \RAMP session as
    $$z \gets \SSC.\keystream(k_\AKE, n).$$
  \item \textbf{Masking.}
    The operation depends on the QKD modes.

  \textit{DV-QKD.}  
  The reconciliation inputs retained by $U$ and $U'$ after the previous steps are binary strings
  $ x,x'\in\{0,1\}^{n}. $
  
  Given a keystream $z$ of length $n$, both users mask by
  \[
    \rkm = x \oplus z,\qquad
    \rkm' = x' \oplus z.
  \]
  It preserves the discrepancy between the paired inputs:
\[
    \rkm\oplus\rkm'
    =
    (x\oplus z)\oplus( x'\oplus z)
    =
    x \oplus x'.
\]
Therefore, the masked results $(\rkm,\rkm')$ neither create new bit errors nor change the locations of existing errors. 
It allows the masked strings to be processed by the same reconciliation procedure as the original inputs.
The masking in DV-QKD can also be directly modeled as an encryption of XOR-Based \SSC as 
\[
\rkm \gets \SSC.\enc(x,z), \quad \rkm' \gets \SSC.\enc(x',z) 
\]
\textit{CV-QKD.} 
In a typical reverse-reconciliation pipeline,
$\usr'$ holds a binary reconciliation string as $x'\in\{0,1\}^n$, while $\usr$ holds
the corresponding log-likelihood ratios (LLRs) as
$x=(L_1,\ldots,L_n)$ for these bits. \RAMP masks this bit-level reconciliation input as
    $\rkm' = x' \oplus z$
and 
$$\widetilde{L}_i=(-1)^{z_i}L_i,\ \forall i \in \{1, \cdots, n\}.$$ 
Specifically, when $z_i=0$, both the bit label and its LLR remain unchanged. 
When $z_i=1$, the bit label is flipped and the corresponding LLR sign is reversed. 
Its magnitude, and therefore its reliability, remains unchanged. We denote these LLRs as $\rkm$.

$(\rkm,\rkm')$ thus represents the same soft-decision decoding problem with a relabeling of binary alphabet.

\end{enumerate}

Thus, for both DV- and
CV-QKD, 
\RAMP hides the effective reconciliation
inputs from service providers while preserving the structure
required by subsequent QKD operations.

\subsection{Post-Processing and Key Distillation}
\label{cons:processing}
\smallskip\noindent 
\textit{$\RKProc()$.}
This stage performs QKD post-processing on the masked raw keys $\rkm$ and $\rkm'$ to eliminate discrepancies and distill the final key $\kfinal$ for $\usr$ and $\usr'$.

\begin{enumerate}

  \item \textbf{Information Reconciliation.} 
  Due to noise and imperfections, 
  $\rkm$ is not identical to $\rkm'$. 
  Users apply error correction
  (\eg, Cascade~\cite{conf/eurocrypt/BrassardS93} in DV-QKD or LDPC~\cite{LDPC} in CV-QKD) to align them with minimal leakage.
  Users conduct correctness verification on reconciled binary strings with the verification procedure of the instantiated QKD. A verification failure causes the session to abort. 
  \item \textbf{Privacy Amplification.} 
  Users compress the reconciled key with a universal hash function to remove information that is potentially leaked. 
  The final key is $\kfinal\in \{0,1\}^{\ell}.$
  
  The output length $\ell$ needs to satisfy both confidentiality branches analyzed in~\cref{sec:security}. In particular, it shall not exceed the secure output length justified by the underlying QKD security analysis, and it also needs to account for all key-dependent public leakage.
\end{enumerate}
After confirmation, $\usr$ (\resp $\usr'$) instructs $\sp$ (\resp $\sp'$) to update $\ST_\sp[\kid]$ (\resp $\ST_{\sp'}[\kid]$)  as \textsf{CONSUMED}. 
If no instruction is received after a predefined period, $\sp$ (\resp $\sp'$) updates $\ST_\sp[\kid]$ (\resp $\ST_{\sp'}[\kid]$) as \textsf{EXPIRED} on its own.

\smallskip\noindent 
\textit{Correctness and Robustness.}
The only modification on raw keys in \RAMP is $\Mask()$, which is realized with \SSC that incurs no error propagation. Consequently, any discrepancies, due to either channel noise or active interference, remain localized. It ensures that the standard QKD post-processing can be conducted identically on the masked raw keys.

As long as the underlying raw keys satisfy the correlation requirements assumed by QKD (\ie, the observed error rate remains below the prescribed threshold), both users derive identical final keys correctly. The property underpins robustness, since the localized errors remain effectively correctable within the tolerance of QKD post-processing. Therefore, \RAMP preserves both correctness and robustness under the same conditions required by conventional QKD.

\smallskip\noindent
\textit{Key-Allocation Integrity.}
The provider signature binds the allocation identifier $\kid$ to the
intended user--provider context, while the provider-side state collection $\ST$ prevents reuse of terminal identifiers. Under EUF-CMA security and
correct state enforcement, a network adversary cannot forge
a fresh authorized allocation or reactivate a consumed identifier.

This mechanism authenticates the allocation context rather than the
contents or physical provenance of the corresponding raw-key material.
Content integrity during acquisition is supplied by the secure
channel for an honest provider endpoint.

\section{Security Formalization}
\label{sec:security}

We formalize the security goals introduced in~\cref{subsec:threat-model}. Our analysis considers a target \RAMP session between two honest users.
The adversary controls the classical network and has the quantum-channel capabilities allowed by QKD.

We target confidentiality with dual-domain resilience, which is captured by the security conditions  $\mathsf{Conf}=\mathbf{M}\vee(\mathbf{Q}\wedge\mathbf{A}).$
Accordingly, we analyze two complementary compromise modes. One considers that both service providers are corrupted while the user-side masking state remains secure. The other allows exposure of the masking state while QKD-service secrecy and independent acquisition remain secure.

\subsection{Provider-Compromise Key Confidentiality} We first consider compromise of the complete QKD-service side. Both service providers may be corrupted and their complete views of the target session may be exposed. 
In this case, the confidentiality relies on condition $\mathsf M$, rather than $\mathsf Q$ or $\mathsf A$.

\begin{definition}[Provider-Compromise Key Confidentiality] 
\label{def:PC}
Let $\Pi$ be a \RAMP scheme. $\mathsf{PC}_{\Pi, \adv}(1^\lambda)$ proceeds as follows.

\begin{enumerate} 
\item The challenger initializes \RAMP and sets a target session between honest users $\usr$ and $\usr'$ served by $\sp$ and $\sp'$. 

\item $\adv$ corrupts $(\sp, \sp')$ and obtains their complete target-session state, including the QKD raw-key materials $(\rk_\sp, \rk_{\sp'})$, metadata $(\md_\sp, \md_{\sp'})$, acquisition credentials, and provider-side protocol state. 
The corrupted providers may deviate arbitrarily from the protocol. 

\item The honest users execute \RAMP. The masking-security condition $\mathbf{M}$ is required to hold. 
In particular, the target \AKE session remains unexposed and $\adv$ does not obtain $k_{\AKE}$ or the resulting keystream $z$. 

\item 
The challenger samples $b\overset{\$}{\leftarrow}\{0,1\}$. 
If the target session aborts, the challenger returns $\bot$, independently of $b$. 
Otherwise, it returns 
$$ K_b = \begin{cases} k_{\mathsf{final}}, & b=0,\\ R\overset{\$}{\leftarrow}\{0,1\}^{\ell}, & b=1. \end{cases} $$ 
together with all public messages of the target session. 

\item The adversary outputs a guess $b'$. \end{enumerate} 

We say $\Pi$ provides provider-compromise final-key confidentiality if for any probabilistic or quantum polynomial-time adversary, the following advantage is negligible in $\lambda$:
$$ \mathsf{Adv}^{\mathsf{PC}}_{\Pi,\adv}(\secparam) = \left| \Pr[b'=b]-\frac{1}{2} \right|. $$

\end{definition}

\cref{thm:pc} asserts provider-compromise confidentiality of \RAMP, with its  proof and quantitative bound in \cref{appx-pc-proof}.
No QKD secrecy or acquisition security is assumed, as the providers may reveal the unmasked QKD-derived input.

\begin{theorem}[Provider-Compromise Confidentiality] 
\label{thm:pc}
Assuming that \AKE provides session-key security, \SSC is \indcpa-secure, and the privacy amplification output length accounts for all key-dependent leakage released after masking, \RAMP provides provider-compromise final-key confidentiality. 
\end{theorem}

\cref{thm:pc} shows that exposing the provider-visible QKD-derived inputs does not compromise the final \RAMP key. The complementary value of QKD will be utilized in \cref{thm:mc}, where the complete masking state may be exposed while confidentiality is retained from the QKD execution.
Hybrid key combiners~\cite{conf/pqcrypto/DowlingHP20,
conf/pqcrypto/BrucknerRS23} provide a related source-failure
property. \RAMP differs in the placement of the QKD service boundary, where
users themselves
execute security-critical processing.

\subsection{Masking-Compromise Key Confidentiality} \label{sec:mc}
We now consider the complementary case in which the complete masking state is exposed. In this experiment, we make no assumption on \AKE or \SSC, but require that  $\mathbf Q$ and $\mathbf A$ hold. 

\begin{definition}[Masking-Compromise Key Confidentiality] 
\label{def:MC}
Let $\Pi$ be a \RAMP scheme. $\mathsf{MC}_{\Pi, \adv}(1^\lambda)$ proceeds as follows. 
\begin{enumerate} 
\item The challenger initializes \RAMP and sets a target session between honest users $\usr$ and $\usr'$ served by $\sp$ and $\sp'$. 

\item The target QKD execution satisfies $\mathbf Q$. Namely, the underlying QKD secrecy and authenticity assumptions hold, while the provider-side state containing the target raw-key materials is not exposed to $\adv$. Whereas $\adv$ could obtain the quantum and classical side information allowed by QKD.

\item Both provider-to-user acquisition paths satisfy $\mathbf A$, so the target raw-key materials and metadata are delivered confidentially and without adversarial modification. 
$\adv$ may still control the public classical network. 

\item The challenger reveals the complete masking state of the target session to $\mathcal A$, including $k_{\mathsf{AKE}}$ and $z$. 

\item The challenger samples $b\overset{\$}{\leftarrow}\{0,1\}$. 
If the target session aborts, the challenger returns $\bot$, independently of $b$. 
Otherwise, it returns 
$$ K_b = \begin{cases} \kfinal, & b=0,\\ R\overset{\$}{\leftarrow}\{0,1\}^{\ell}, & b=1. \end{cases} $$ 
together with all public messages of the target session. 
\item $\mathcal A$ outputs a guess $b'$. 
\end{enumerate} 

We say $\Pi$ provides masking-compromise final-key confidentiality if for any probabilistic or quantum polynomial-time adversary, the following advantage is negligible in $\lambda$: 
$$ \mathsf{Adv}^{\mathsf{MC}}_{\Pi, \adv}(\secparam) = \left| \Pr[b'=b]-\frac{1}{2} \right|. $$ 
\end{definition}

\cref{thm:mc} asserts masking-compromise confidentiality of \RAMP,
with its proof and quantitative bound in \cref{appx-mc-proof}.
It makes no assumption on the secrecy of \AKE or \SSC, as the target
masking state, including $k_{\mathsf{AKE}}$ and $z$, may be
exposed.

\begin{theorem}[Masking-Compromise Confidentiality]
\label{thm:mc}
Assuming that $\mathsf Q$ and $\mathsf A$ hold, the two acquisition
channels provide secure-channel security, the underlying QKD protocol provides $\varepsilon_{\mathsf{qkd}}$-secrecy, and the instantiated
reconciliation procedure satisfies public-mask invariance (\cref{lem:invar}),
\RAMP provides masking-compromise final-key confidentiality.
\end{theorem}

Based on~\cref{thm:pc,thm:mc}, \cref{appx:dual-domain} derives a corollary on dual-domain compromise resilience of \RAMP.

\subsection{Detectability of Inconsistent Key Material}
Detectability in \RAMP refers to protocol abortion when the supplied
raw-key materials violate the admissibility or correctness conditions
of the instantiated QKD post-processing procedure. Such inconsistency
may be detected during parameter estimation, reconciliation, or
correctness verification.

\begin{theorem}[Detectability of Inconsistent Key Material]
\label{thm:detectability}
Assume that the underlying QKD post-processing procedure accepts
inconsistent or otherwise inadmissible input material with probability
at most $\varepsilon_{\mathsf{det}}$. If any one of the service providers supplies
such material, then RAMP accepts the resulting session with probability
at most $\varepsilon_{\mathsf{det}}$.
\end{theorem}

\RAMP performs parameter estimation before masking, so its
acceptance behavior remains. For subsequent processing, the masking transformation preserves the discrepancy or soft-decision structure of the reconciliation problem, and hence preserves its
success or failure behavior. 
The same relabeling also preserves whether the reconciled binary strings agree, so correctness verification retains its underlying detection guarantee. Therefore,
\RAMP preserves the relevant acceptance and rejection behavior of the underlying QKD post-processing. \cref{thm:detectability} directly follows from
the QKD soundness.

The unilateral detectability also brings a useful \emph{self-checking feature}. 
Suppose the two providers are independently administered and at most one deviates. 
A deviating provider may degrade the correlation of its supplied material, \eg, by perturbing its local outcomes. 
However, it cannot deliberately improve its correlation with the honest-side undisclosed samples beyond the correlation already available from its own QKD view, because those samples are unknown when the raw-key blocks are acquired. 
In \RAMP, the user-side parameter estimation over the paired material provides a direct consistency check on their correlation structure.

\section{Experimental Evaluation}
\label{sec:real_qkd_eval}

\subsection{Experimental Setup and Scope}
Our experiment evaluates \RAMP from two aspects. 
\cref{exp:user-side-processing,exp:reconciliation} evaluate whether real QKD material can be processed through the masked reconciliation and privacy-amplification pipeline on general-purpose end-user hardware. 
\cref{exp:ramp} quantifies the system overhead introduced specifically by \RAMP, including raw-key acquisition, session binding, post-quantum provider authentication, and protocol metadata.
Throughout the evaluation, we employ experimentally acquired QKD data from qOptica~100 CV-QKD testbed~\cite{qoptica100}.

\smallskip\noindent 
\textit{Platforms.}
To characterize the overhead across hardware an end user could realistically own, we use two laptops of different capabilities without GPU acceleration. 
Specifically, we run experiments on a \emph{high-end} laptop as Apple MacBook Pro (M4~Pro, 24\, GB RAM) and an \emph{ordinary} laptop, a Dell Latitude 7430 (Intel Core i7-1265U, 16 GB RAM), respectively.

\smallskip\noindent 
\textit{Post-Processing Implementation.}
We prototype this user-side pipeline in MATLAB, a common choice compatible with QKD implementation. 
Taking as input the raw-key materials from service providers and the masking key established by users, the pipeline conducts parameter estimation, masking, information reconciliation, and privacy amplification to distill the final key of the \RAMP session.
We instantiate \SSC for keystream generation with AES-256-CTR.
We employ forward error correction with LDPC codes of block length $n_H = 2\times10^6$ bits and rate $R=1/10$.
We aggregate $200$ LDPC blocks, which correspond to $4\times10^8$ accepted bits, before privacy amplification.
\cref{appx:config} details the configuration.

\smallskip\noindent 
\textit{\RAMP Control-Plane Implementation.}
We prototype \RAMP's setup, registration, raw-key acquisition, session binding, and authenticated masking-key establishment in Python. On both platforms, we evaluate a paired raw-key workload derived from CV-QKD with a size of approximately $3.65$GB in total. 

We instantiate provider signatures \SIG with NIST-standardized ML-DSA-65~\cite{ml-dsa}. 
The acquisition channels are instantiated with TLS~1.3 in PSK mode~\cite{rfc9846,rfc9257}, using independently provisioned acquisition credentials. The user-to-user \AKE is instantiated with mutually authenticated TLS 1.3 from OpenSSL~\cite{openssl} with NIST-standardized ML-KEM-768~\cite{ml-kem} for key establishment, after which a $32$-byte session key $k_\AKE$ is exported and bound to the \RAMP session.

\subsection{User-Side QKD Processing}
\label{exp:user-side-processing}

\begin{table}[t]
\centering
\caption{User-Side Processing Time (s) on $200$ LDPC Blocks of Length $n_H=2\times10^6$ ($4\times10^8$ accepted bits)
}
\label{tab:exp-breakdown}
\begin{tabular}{l r r}
\toprule
Operation & M4~Pro & i7-1265U \\
\midrule
\multicolumn{3}{l}{\textit{Shared / Public (both $\usr, \usr'$)}}\\
\quad Load and clean data            & 127.276 & 34.110 \\
\quad Rescale and sampling           & 3.723   & 14.257 \\
\quad Key split                      & 2.600   & 19.588 \\
\quad Parameter estimation           & 0.318   & 0.561 \\
\quad Masking keystream              & 46.759 & 248.425 \\
\midrule
\multicolumn{3}{l}{\textit{$\usr$ (soft info $+$ decode $+$ PA)}}\\
\quad Octonion map $+$ LLRs          & 142.514 & 616.830 \\
\quad Mask apply (LLR sign flip)     & 0.198   & 0.421 \\
\quad LDPC decode ($200$ blocks)     & 717.011 & 4165.207 \\
\quad Verification tag               & 14.664  & 45.770 \\
\quad Privacy amplification          & 43.047  & 155.902 \\
\midrule
\multicolumn{3}{l}{\textit{$\usr'$ (lightweight)}}\\
\quad Mask apply (bit XOR)           & 0.080   & 0.137 \\
\quad Syndrome ($200$ blocks)        & 2.829 & 18.043 \\
\quad Verification tag               & 15.070  & 45.755 \\
\midrule
\quad $\usr$ total                   & 1098.111 & 5301.071 \\
\quad $\usr'$ total                  & 152.270  & 132.824 \\
\quad \textbf{Wall-clock total}      & \textbf{1123.514} & \textbf{5387.427} \\
\bottomrule
\end{tabular}
\end{table}

\cref{tab:exp-breakdown} reports the computational cost when aggregating $200$ LDPC blocks ($4\times10^8$ accepted bits). 
It comfortably exceeds the finite-size threshold, which we evaluate in~\cref{exp:reconciliation}. 

\smallskip\noindent 
\textit{Dominant Reconciliation vs. Lightweight Masking.}
LDPC decoding is the dominant computational stage on both platforms. It requires $717.0$s on M4 Pro and $4165.2$s on i7-1265U, corresponding to $3.6$s and $20.8$s per block, respectively.
The complete finite-size processing run finishes in approximately $18.7$min on M4 Pro and $89.8$min on  i7-1265U.

Importantly, this dominant cost is inherited from standard QKD processing rather than introduced by the \RAMP service. The \RAMP-specific masking work consists of keystream generation and the two local mask applications.
These operations take approximately $47.0$s on M4 Pro and $249.0$s on i7-1265U, accounting for only approximately $4.19\%$ and $4.62\%$ of the respective wall-clock runtimes.
It confirms that masking for hiding keying inputs only incurs a small incremental cost.

\smallskip\noindent 
\textit{Asymmetric Workload.}
\RAMP also preserves the inherent asymmetry of syndrome-based reverse reconciliation.
$\usr'$ performs only lightweight syndrome computation, essentially a sparse matrix-vector product, with $14.1$ms (\resp $90.2$ms) per block on the M4~Pro (\resp i7-1265U).
By contrast, $\usr$ bears the decode-heavy reconciliation. 
Consequently the $\usr'$ total time is nearly independent of platform capability ($152.3$s vs.\ $132.8$s), while the $\usr$ total time differs by almost $5\times$. 

This distinction is useful in heterogeneous deployments. RAMP does not require both users to possess equivalent computational resources.
More capable users handle the computationally intensive decoding, while resource-constrained devices work on the opposite side with minimal local overhead. 
\cref{appx:findings} details its effect on deployment and optimization.

\subsection{Reconciliation Efficiency and Key Yield}
\label{exp:reconciliation}
Parameter estimation on the acquired QKD data yields an operating point with $\mathrm{SNR} = 0.161$ and a channel capacity
\[
C_{AB}=\log_2(1+\mathrm{SNR}) \approx 0.215~\text{bits/use}.
\]
``bits/use'' denotes secret bits per quantum channel use, or equivalently per transmitted signal state.
For the LDPC code with a rate $R=1/10$, the effective reconciliation efficiency is
$$\beta_{\mathrm{eff}}
=\frac{2R}{C_{AB}}\approx 0.929,$$
where the factor of $2$ accounts for the two quadratures obtained by heterodyne detection. We observe zero frame errors and zero residual post-reconciliation bit errors over the evaluated blocks.
Thus, at the evaluated operating point, masking in \RAMP remains compatible with high-efficiency reconciliation without changing the observed decoding outcome.

The estimated channel model gives a Holevo bound on the eavesdropper's information of
$$\chi_{E|B} \approx 0.182~\text{bits/use},$$
which is computed under the collective-Gaussian-attack model adopted by our CV-QKD testbed~\cite{qoptica100}.

The corresponding asymptotic key rate is
$$K_{\infty}=\beta_{\mathrm{eff}}C_{AB}-\chi_{E|B}=2R-\chi_{E|B}\approx 0.018~\text{bits/use},$$
confirming operation in a positive-key regime. 
The channel parameters are estimated jointly from both quadratures over the parameter-estimation sample budget.

\smallskip\noindent
\textit{Finite-Size Effects.}
Finite-size analysis introduces a privacy-amplification penalty $\delta(N)$ that decreases with the aggregated raw-key length $N$, reducing the secret fraction to 
$$r_{\mathrm{sec}}(N) = 2R - \chi_{E|B} - \delta(N).$$
A positive finite-size key therefore requires accumulating enough blocks to drive $\delta(N)$ below the asymptotic margin $K_{\infty}\approx 0.018$. With our security parameters this threshold is
\[
N_{\min} \approx 2\times10^8~\text{bits}~(\approx 100~\text{LDPC blocks}).
\]
It is a modest requirement that follows directly from the key-rate margin with the both-quadrature parameter estimation.

For the aggregated run of $N = 4\times10^8$ bits ($200$ blocks) reported in \cref{tab:exp-breakdown}, we obtain $\delta(N)\approx 0.0126$ and hence a finite-size secret fraction $r_{\mathrm{sec}}\approx 0.0057$~bits/use, yielding a final key of $\ell \approx 2.28\times10^6$ bits after privacy amplification. Privacy amplification over this $4\times10^8$-bit input completes in $43.0$s on M4~Pro (\resp $155.9$s on i7-1265U) using the FFT-based Toeplitz extractor.
It confirms that privacy amplification is not a bottleneck, as it is a small fraction of the decode time.

The effective key-distillation throughputs are around
\[
2.1\times10^3~\text{bits/s}~\text{(M4~Pro)}, \quad
4.2\times10^2~\text{bits/s}~\text{(i7-1265U)}.
\]
Viewed as $256$-bit symmetric-key material, these rates correspond to approximately $8.2$ and $1.6$ AES-256 key-equivalents per second, which are \emph{derived from quantum entropy}.
The result demonstrates the computational throughput of the complete user-side post-processing pipeline at the evaluated finite-size operating point. 
The acquisition and masking-key establishment are measured separately in~\cref{exp:ramp}.
Privacy amplification remains subdominant to LDPC decoding.

\smallskip\noindent
\textit{On Aggregated Block Count.}
The $200$-block result exhibits the complete user-side QKD post-processing workload at a finite-size-secure operating point ($N > N_{\min}$). The asymptotic rate, the Holevo bound, and the finite-size threshold above are all properties of the estimated channel and are independent of how many blocks are processed. Certifying a fully independent finite-size key of a specific length additionally requires that the aggregated input consist of independent raw-key material of the corresponding size, which is a function of the acquired data volume rather than of the processing pipeline.

\subsection{\RAMP Service-Path Evaluation}
\label{exp:ramp}
The evaluation follows the execution order of \RAMP. It allows us to clarify one-time provisioning costs and per-session operations, identifying where the system spends time along an actual execution.
Online operations are measured over $30$ executions after $5$ warm-up runs, cryptographic primitives over $100$ executions, and one-time setup over $10$ executions.

\subsubsection{Setup and Raw-Key Provisioning}
\RAMP begins with provider initialization and QKD raw-key generation. We implement provider initialization and the storage update associated with importing already generated QKD material. The physical QKD 
stage itself is due to qOptica~100 CV-QKD testbed~\cite{qoptica100} and independent of our evaluation.

The identity initialization of providers (and users) is complete in the tens-of-milliseconds range. Importing and indexing the prepared $3.65$GB paired raw-key workload takes $1.32$s on M4 Pro and $2.68$s on i7-1265U, including integrity hashing, identifier derivation, and provider-side storage updates.

\subsubsection{Registration and Raw-Key Acquisition}
\smallskip\noindent
\textit{Registration.}
$\Regist()$
provisions the independent symmetric acquisition credential $k_{\mathsf{acq}}$ before raw-key acquisition. Its cost is only sub-millisecond to several milliseconds depending on both platforms, minor compared with subsequent acquisition.

\begin{table*}[t]
\centering
\caption{\RAMP Service-Path Performance
}
\label{tab:exp-ramp}
\begin{tabular}{l l r r}
\toprule
Protocol Stage & Measured Operation & M4~Pro & i7-1265U \\
\midrule
Setup (\cref{cons:setup}) &  Raw-Key provisioning            & $1.32$s & $2.68$s \\
\midrule
\multirow{4}{*}{Acquisition (\cref{cons:register})} 
& Raw-key acquisition            & $13.99$s & $22.95$s \\
& Acquisition throughput           & $260.6$MB/s   & $158.8$MB/s \\
& ML-DSA-65 Sign           & $7.48$ms  & $5.53$ms  \\
& ML-DSA-65 Verify              & $6.82$ms & $4.39$ms \\
\midrule
Session binding (\cref{sec:mask}) & $\sid$ computation               & $19.1\mu$s   & $24.3\mu$s \\
\multirow{2}{*}{Masking-key establishment (\cref{sec:mask})} & 
ML-KEM-768 $\AKE^\dag$ time & $2.33$s   & $2.28$s \\
 & ML-KEM-768 $\AKE$ success rate & $30/30$   & $30/30$ \\
\bottomrule
\end{tabular}
\begin{tablenotes}
    \item $\dag$ It is the wall-clock time of the current implementation wrapper, including OpenSSL subprocess management. 
\end{tablenotes}
\end{table*}

\smallskip\noindent
\textit{Raw-Key Acquisition.}
We then execute $\RKAcq()$ over the $3.65$GB paired workload. The measured path includes provider-side raw-key retrieval and integrity checking, serialization, TLS~1.3 PSK protection under the independently provisioned acquisition credentials, local transport, deserialization, and persistence at the two users.

The Mac completes acquisition in a median of $13.99$s, corresponding to $260.6$MB/s of application-level throughput. The Dell requires $22.95$s, corresponding to $158.8$MB/s. Across the $30$ measured runs, the interquartile latency ranges are $13.45$--$14.32$s and $21.46$--$24.52$s, respectively.
The result characterizes the protected acquisition path rather than a pure network or cryptographic primitive.

We specifically evaluate the time of authenticating the binding between the selected $\kid$, users, and providers using ML-DSA-65.
Median signing and verification costs are $7.48$ms and $6.82$ms on M4 Pro, and $5.53$ms and $4.39$ms on i7-1265U, respectively. A signing-and-verification pair therefore accounts for only approximately $0.10$\% and $0.04$\% of the respective acquisition times. 
Thus, provider authentication is not the dominant cost of $\RKAcq()$. The acquisition cost mainly comes from the transmission of the raw-key material.

\subsubsection{Session Binding and Masking-Key Establishment}
\smallskip\noindent
\textit{Session Binding.}
After acquisition, the two users bind the session by computing the hash of $(U, U', P, P', \kid)$ as $\sid$. 
The operation takes a median of only $19.1\mu$s on M4 Pro and $24.3\mu$s on i7-1265U, which is minor relative to both raw-key acquisition and subsequent QKD processing.

\smallskip\noindent
\textit{Pre-Mask QKD Preparation.}
Consistent with the construction in~\cref{sec:mask}, reconciliation inputs are fixed before the masking key is established. This operation belongs to the QKD-processing pipeline rather than the RAMP control-plane, therefore evaluated together with multidimensional reconciliation and post-processing in~\cref{exp:user-side-processing}.

\smallskip\noindent
\textit{Masking-Key Establishment.}
Once the reconciliation inputs and $\sid$ are fixed, the users execute the authenticated key-establishment step. Across all $60$ measured sessions ($30$ per platform), the implementation successfully negotiates the ML-KEM-768 TLS group and derives a fresh 32-byte $k_\AKE$.

The current implementation completes the masking-key-establishment wrapper in a median of $2.33$s on M4 Pro and $2.28$s on i7-1265U. It is executed once per \RAMP session. However, the prototype invokes an external OpenSSL server process~\cite{openssl}, and the measured wall-clock interval also includes process startup, synchronization, and teardown. 
The quantity reports an implementation-level masking-key-establishment cost, rather than a primitive-level benchmark.

\subsection{Evaluation Summary}
\smallskip\noindent
\textit{\RAMP preserves practical QKD post-processing.}
On experimentally acquired CV-QKD data, the pipeline achieves a reconciliation efficiency of $\beta_{\mathrm{eff}}\approx 0.929$, with no observed frame errors or residual post-reconciliation bit errors. 
The complete user-side prototype processes this workload at approximately $2.1$Kb/s on the M4 Pro and $0.42$Kb/s on the i7-1265U.

\smallskip\noindent
\textit{\RAMP-specific protection remains a small fraction of the overall cost.}
Mask generation and application account for only $4.19\%$ and $4.62\%$ of the measured post-processing wall-clock time on the two platforms, while session binding and ML-DSA authentication remain in the microsecond-to-millisecond regime. 
The current masking-key-establishment wrapper adds approximately $2.3$s per session, including OpenSSL process-management overhead.
The user-side QKD post-processing takes $1123.5$s and $5387.4$s. The measured cost is therefore dominated by the reconciliation of the QKD material rather than by the \RAMP service path.

\smallskip\noindent
\textit{The experiment reveals a valuable separation between service provisioning and user-side computation.}
The LDPC reconciliation is performed after acquisition at the users. As per our results, this workload is itself asymmetric, \ie, $\usr$ executes the decode-heavy operation, whereas $\usr'$ performs lightweight syndrome generation. The protocol therefore executes more efficiently across two devices of substantially different capabilities, still requiring no QKD hardware at either endpoint.

Overall, the experiments establish \emph{feasibility rather than a production-performance bound}. 
Our evaluation is on one low-SNR CV-QKD operating point, one LDPC configuration, and a local acquisition path. Different QKD parameters, optimized decoders, networks, and hardware will affect performance. 
Despite the boundaries, our result shows that \RAMP successfully moves QKD post-processing to commodity user devices with modest control costs.  
\RAMP enables users without QKD hardware to locally distill QKD-derived keys.

\section{Conclusion}
\label{sec:conclusion}
\RAMP elevates QKD from a device-centric model to a \emph{service-oriented architecture}. 
By decoupling quantum-state generation from user-side processing, \RAMP enables ordinary users to obtain quantum-derived keys while providing final-key confidentiality under complementary compromise modes.

\bibliographystyle{plainurl}
\bibliography{sample}

@inproceedings{conf/ispec/KozlovicsPLV23,
  author       = {Sergejs Kozlovics and
                  Krisjanis Petrucena and
                  Davis Larins and
                  Juris Viksna},
  title        = {Quantum Key Distribution as a Service and Its Injection into {TLS}},
  booktitle    = {{ISPEC}},
  pages        = {527--545},
  year         = {2023}
}

@inproceedings{focs/Shor94,
  author       = {Peter W. Shor},
  title        = {Algorithms for Quantum Computation: Discrete Logarithms and Factoring},
  booktitle    = {{FOCS}},
  pages        = {124--134},
  year         = {1994}
}

@article{leverrier2008multidimensional,
  title={Multidimensional reconciliation for continuous-variable quantum key distribution},
  author={Leverrier, Anthony and Alléaume, Romain and Boutros, Joseph and Z{\'e}mor, Gilles and Grangier, Philippe},
  journal={Physical Review A},
  volume={77},
  number={4},
  pages={042325},
  year={2008}
}

@article{jouguet2011long,
  title={Long-distance continuous-variable quantum key distribution with a Gaussian modulation},
  author={Jouguet, Paul and Kunz-Jacques, Simon and Leverrier, Anthony and Grangier, Philippe and Diamanti, Eleni},
  journal={Physical Review A},
  volume={84},
  number={6},
  pages={062317},
  year={2011}
}

@misc{qoptica100,
  title        = {qOptica 100 CV-QKD System},
  author       = {{Quintessence Labs}},
  howpublished = {\url{https://www.quintessencelabs.com}},
  note         = {Product specification sheet}
}

@book{books/crc/MenezesOV96,
  author       = {Alfred Menezes and
                  Paul C. van Oorschot and
                  Scott A. Vanstone},
  title        = {Handbook of Applied Cryptography},
  publisher    = {{CRC} Press},
  year         = {1996}
}

@article{journals/entropy/LorunserKPS23,
  author       = {Thomas Lor{\"{u}}nser and
                  Stephan Krenn and
                  Christoph Pacher and
                  Bernhard Schrenk},
  title        = {On the Security of Offloading Post-Processing for Quantum Key Distribution},
  journal      = {Entropy},
  volume       = {25},
  number       = {2},
  pages        = {226},
  year         = {2023}
}

@article{journals/cee/RamalingamRKL24,
  author       = {Jothi Ramalingam and
                  Satyabrat Rath and
                  Lakshmi Kuppusamy and
                  Cheng{-}Chi Lee},
  title        = {Accelerating {QKD} post-processing by secure offloading of information
                  reconciliation},
  journal      = {Comput. Electr. Eng.},
  volume       = {120},
  pages        = {109721},
  year         = {2024}
}

@misc{chennistir,
  title={{IR 8105}, Report on Post-Quantum Cryptography},
  author={NIST},  
  howpublished = {\url{https://csrc.nist.gov/pubs/ir/8105/final}},
  year={2016}
}

@misc{ETSI,
  title={Quantum-Safe Cryptography (QSC); Impact of Quantum Computing
on Symmetric Cryptography},
  author={{ETSI TR 103 967}},
  year={2025}
}

@misc{ETSIAPI,
  title={Quantum Key Distribution (QKD); Protocol and data format of REST-based key delivery API},
  author={{ETSI GS QKD 014}},
  year={2019}
}

@article{journals/tcs/BennettB14,
  author       = {Charles H. Bennett and
                  Gilles Brassard},
  title        = {Quantum cryptography: Public key distribution and coin tossing},
  journal      = {Theor. Comput. Sci.},
  volume       = {560},
  pages        = {7--11},
  year         = {2014},
  note         = {(conference version published in {\it IEEE Int. Conf. Comput., Syst. Signal}, 1984)}
}

@article{wootters1982single,
  title={A single quantum cannot be cloned},
  author={Wootters, William K and Zurek, Wojciech H},
  journal={Nature},
  volume={299},
  number={5886},
  pages={802--803},
  year={1982}
}

@article{ralph1999continuous,
  title={Continuous variable quantum cryptography},
  author={Ralph, Timothy C},
  journal={Physical Review A},
  volume={61},
  number={1},
  pages={010303},
  year={1999},
  publisher={APS}
}

@inproceedings{stoc/Grover96,
  author       = {Lov K. Grover},
  title        = {A Fast Quantum Mechanical Algorithm for Database Search},
  booktitle    = {{STOC}},
  pages        = {212--219},
  publisher    = {{ACM}},
  year         = {1996}
}

@inproceedings{conf/pqcrypto/AnandTTU16,
  author       = {Mayuresh Vivekanand Anand and
                  Ehsan Ebrahimi Targhi and
                  Gelo Noel Tabia and
                  Dominique Unruh},
  title        = {Post-Quantum Security of the CBC, CFB, OFB, CTR, and {XTS} Modes of
                  Operation},
  booktitle    = {PQCrypto},
  pages        = {44--63},
  year         = {2016}
}

@misc{ml-kem,
  title = {Module-Lattice-Based Key-Encapsulation Mechanism Standard},
  author={NIST},
  howpublished = {\url{https://doi.org/10.6028/NIST.FIPS.203}},
  year = 2024
}

@misc{ml-dsa,
  title = {Module-Lattice-Based Digital Signature Standard},
  author={NIST},
  year = 2024,
  howpublished = {\url{https://doi.org/10.6028/NIST.FIPS.204}},
}

@article{LangloisS15,
  author       = {Adeline Langlois and
                  Damien Stehl{\'{e}}},
  title        = {Worst-case to average-case reductions for module lattices},
  journal      = {Des. Codes Cryptogr.},
  volume       = {75},
  number       = {3},
  pages        = {565--599},
  year         = {2015}
}

@inproceedings{ndss/SikeridisKD20,
  author       = {Dimitrios Sikeridis and
                  Panos Kampanakis and
                  Michael Devetsikiotis},
  title        = {Post-Quantum Authentication in {TLS} 1.3: {A} Performance Study},
  booktitle    = {{NDSS}},
  year         = {2020}
}

@inproceedings{pqcrypto/BindelHMS17,
  author       = {Nina Bindel and
                  Udyani Herath and
                  Matthew McKague and
                  Douglas Stebila},
  title        = {Transitioning to a Quantum-Resistant Public Key Infrastructure},
  booktitle    = {{PQCrypto}},
  pages        = {384--405},
  year         = {2017}
}

@misc{openssl,
  author       = {{The OpenSSL Project}},
  title        = {{OpenSSL: The Open Source Toolkit for SSL/TLS}},
  howpublished = {\url{https://www.openssl.org/}},
  year         = {2025},
  note         = {Version 3.5}
}

@inproceedings{conf/pqcrypto/BrucknerRS23,
  author       = {Sonja Bruckner and
                  Sebastian Ramacher and
                  Christoph Striecks},
  title        = {Muckle+: End-to-End Hybrid Authenticated Key Exchanges},
  booktitle    = {{PQCrypto}},
  pages        = {601--633},
  year         = {2023}
}

@inproceedings{conf/pqcrypto/DowlingHP20,
  author       = {Benjamin Dowling and
                  Torben Brandt Hansen and
                  Kenneth G. Paterson},
  title        = {Many a Mickle Makes a Muckle: {A} Framework for Provably Quantum-Secure
                  Hybrid Key Exchange},
  booktitle    = {{PQCrypto}},
  pages        = {483--502},
  year         = {2020}
}

@inproceedings{conf/eurocrypt/BrassardS93,
  author       = {Gilles Brassard and
                  Louis Salvail},
  title        = {Secret-Key Reconciliation by Public Discussion},
  booktitle    = {{EUROCRYPT}},
  pages        = {410--423},
  year         = {1993}
}

@article{LDPC,
author = {Elkouss, David and Martinez-mateo, Jesus and Martin, Vicente},
title = {Information reconciliation for quantum key distribution},
year = {2011},
volume = {11},
number = {3},
issn = {1533-7146},
journal = {Quantum Info. Comput.},
month = mar,
pages = {226–238},
numpages = {13}
}

@article{bernstein2017post,
  title={Post-quantum cryptography},
  author={Bernstein, Daniel J and Lange, Tanja},
  journal={Nature},
  volume={549},
  number={7671},
  pages={188--194},
  year={2017}
}

@article{mehic2020quantum,
  title={Quantum key distribution: a networking perspective},
  author={Mehic, Miralem and Niemiec, Marcin and Rass, Stefan and Ma, Jiajun and Peev, Momtchil and Aguado, Alejandro and Martin, Vicente and Schauer, Stefan and Poppe, Andreas and Pacher, Christoph and others},
  journal={ACM Computing Surveys (CSUR)},
  volume={53},
  number={5},
  pages={1--41},
  year={2020}
}

@article{sasaki2011field,
  title={Field test of quantum key distribution in the Tokyo QKD Network},
  author={Sasaki, Masahide and Fujiwara, Mikio and Ishizuka, H and Klaus, W and Wakui, K and Takeoka, M and Miki, S and Yamashita, T and Wang, Z and Tanaka, A and others},
  journal={Optics express},
  volume={19},
  number={11},
  pages={10387--10409},
  year={2011},
  publisher={Optical Society of America}
}

@article{journals/comsur/CaoZWZNH22,
  author       = {Yuan Cao and
                  Yongli Zhao and
                  Qin Wang and
                  Jie Zhang and
                  Soon Xin Ng and
                  Lajos Hanzo},
  title        = {The Evolution of Quantum Key Distribution Networks: On the Road to
                  the Qinternet},
  journal      = {{IEEE} Commun. Surv. Tutorials},
  volume       = {24},
  number       = {2},
  pages        = {839--894},
  year         = {2022}
}

@article{peev2009secoqc,
  title={The SECOQC quantum key distribution network in Vienna},
  author={Peev, Momtchil and Pacher, Christoph and All{\'e}aume, Romain and Barreiro, Claudio and Bouda, Jan and Boxleitner, Winfried and Debuisschert, Thierry and Diamanti, Eleni and Dianati, Mehrdad and Dynes, James F and others},
  journal={New journal of physics},
  volume={11},
  number={7},
  pages={075001},
  year={2009},
  publisher={IOP Publishing}
}

@article{stucki2011long,
  title={Long-term performance of the SwissQuantum quantum key distribution network in a field environment},
  author={Stucki, Damien and Legre, Matthieu and Buntschu, Francois and Clausen, B and Felber, Nadine and Gisin, Nicolas and Henzen, Luca and Junod, Pascal and Litzistorf, G{\'e}rald and Monbaron, Patrick and others},
  journal={New Journal of Physics},
  volume={13},
  number={12},
  pages={123001},
  year={2011},
  publisher={IOP Publishing}
}

@article{inoue2002differential,
  title={Differential phase shift quantum key distribution},
  author={Inoue, Kyo and Waks, Edo and Yamamoto, Yoshihisa},
  journal={Physical review letters},
  volume={89},
  number={3},
  pages={037902},
  year={2002}
}

@article{xu2020secure,
  title={Secure quantum key distribution with realistic devices},
  author={Xu, Feihu and Ma, Xiongfeng and Zhang, Qiang and Lo, Hoi-Kwong and Pan, Jian-Wei},
  journal={Reviews of Modern Physics},
  volume={92},
  number={2},
  pages={025002},
  year={2020},
  publisher={APS}
}

@article{liao2018satellite,
  title={Satellite-relayed intercontinental quantum network},
  author={Liao, Sheng-Kai and Cai, Wen-Qi and Handsteiner, Johannes and Liu, Bo and Yin, Juan and Zhang, Liang and Rauch, Dominik and Fink, Matthias and Ren, Ji-Gang and Liu, Wei-Yue and others},
  journal={Physical review letters},
  volume={120},
  number={3},
  pages={030501},
  year={2018},
  publisher={APS}
}

@article{cao2019sdqaas,
  title={SDQaaS: Software defined networking for quantum key distribution as a service},
  author={Cao, Yuan and Zhao, Yongli and Wang, Jianquan and Yu, Xiaosong and Ma, Zhangchao and Zhang, Jie},
  journal={Optics express},
  volume={27},
  number={5},
  pages={6892--6909},
  year={2019},
  publisher={Optical Society of America}
}

@article{cao2019multi,
  title={Multi-tenant secret-key assignment over quantum key distribution networks},
  author={Cao, Yuan and Zhao, Yongli and Lin, Rui and Yu, Xiaosong and Zhang, Jie and Chen, Jiajia},
  journal={Optics Express},
  volume={27},
  number={3},
  pages={2544--2561},
  year={2019},
  publisher={Optical Society of America}
}

@book{books/crc/KatzLindell2014,
  author       = {Jonathan Katz and
                  Yehuda Lindell},
  title        = {Introduction to Modern Cryptography, Second Edition},
  publisher    = {{CRC} Press},
  year         = {2014},
}

@misc{journals/rfc/rfc6063,
    series =    {Request for Comments},
    number =    6063,
    howpublished =  {RFC 6063},
    publisher = {RFC Editor},
    author =    {A. Doherty and M. Pei and S. Machani and M. Nystrom},
    title =     {{Dynamic Symmetric Key Provisioning Protocol (DSKPP)}},
    pagetotal = 105,
    year =      2010,
    month =     dec
}

@article{journals/corr/PortmannR14,
  author       = {Christopher Portmann and
                  Renato Renner},
  title        = {Cryptographic security of quantum key distribution},
  journal      = {CoRR},
  volume       = {abs/1409.3525},
  year         = {2014}
}

@article{journals/quantum/TomamichelL17,
  author       = {Marco Tomamichel and
                  Anthony Leverrier},
  title        = {A largely self-contained and complete security proof for quantum key
                  distribution},
  journal      = {Quantum},
  volume       = {1},
  pages        = {14},
  year         = {2017}
}

@article{journals/jsac/GarciaASORM25,
  author       = {Carlos Rubio Garcia and
                  Abraham Cano Aguilera and
                  Catalina Ioana Stan and
                  Juan Jos{\'{e}} Vegas Olmos and
                  Simon Rommel and
                  Idelfonso Tafur Monroy},
  title        = {Enhanced Network Security Protocols for the Quantum Era: Combining
                  Classical and Post-Quantum Cryptography, and Quantum Key Distribution},
  journal      = {{IEEE} J. Sel. Areas Commun.},
  volume       = {43},
  number       = {8},
  pages        = {2765--2781},
  year         = {2025}
}

@article{milicevic2018quasi,
  title={Quasi-cyclic multi-edge LDPC codes for long-distance quantum cryptography},
  author={Mario Milicevic and Chen Feng and Lei M. Zhang and P. Glenn Gulak},
  journal={npj Quantum Information},
  volume={4},
  number={1},
  pages={21},
  year={2018},
  publisher={Nature Publishing Group UK London}
}

@article{PhysRevA.109.012604,
  title = {High-efficiency rate-adaptive reconciliation in continuous-variable quantum key distribution},
  author = {Hongzhao Yang and Shuaishuai Liu and Shenshen  Yang and Zhenguo Lu and Yanxia Li and Yongmin Li},
  journal = {Phys. Rev. A},
  volume = {109},
  issue = {1},
  pages = {012604},
  numpages = {8},
  year = {2024}
}

@article{yin2017satellite,
  title={Satellite-based entanglement distribution over 1200 kilometers},
  author = {Juan Yin  and Yuan Cao  and Yuhuai Li  and Shengkai Liao  and Liang Zhang  and Jigang Ren  and Wenqi Cai  and Weiyue Liu  and Bo Li  and Hui Dai  and Guangbing Li  and Qiming Lu  and Yunhong Gong  and Yu Xu  and Shuanglin Li  and Fengzhi Li  and Yayun Yin  and Ziqing Jiang  and Ming Li  and Jianjun Jia  and Ge Ren  and Dong He  and Yilin Zhou  and Xiaoxiang Zhang  and Na Wang  and Xiang Chang  and Zhencai Zhu  and Naile Liu  and Yuao Chen  and Chaoyang Lu  and Rong Shu  and Chengzhi Peng  and Jianyu Wang  and Jianwei Pan },
  journal={Science},
  volume={356},
  number={6343},
  pages={1140--1144},
  year={2017}
}

@inproceedings{Stanley_2022,
  author = {M. Stanley and Y. Gui and D. Unnikrishnan and S.R.G. Hall  and I. Fatadin},
  title = {Recent Progress in Quantum Key Distribution Network Deployments and Standards},
  booktitle={Journal of Physics: Conference Series},
  volume={2416},
  pages={012001},
  year={2022},
  organization={IOP Publishing}
}

@article{PhysRevA.72.012332,
  title = {Information-theoretic security proof for quantum-key-distribution protocols},
  author = {Renner, Renato and Gisin, Nicolas and Kraus, Barbara},
  journal = {Phys. Rev. A},
  volume = {72},
  issue = {1},
  pages = {012332},
  numpages = {17},
  year = {2005}
}

@inproceedings{conf/eurocrypt/CanettiK01,
  author       = {Ran Canetti and
                  Hugo Krawczyk},
  editor       = {Birgit Pfitzmann},
  title        = {Analysis of Key-Exchange Protocols and Their Use for Building Secure
                  Channels},
  booktitle    = {{EUROCRYPT}},
  series       = {Lecture Notes in Computer Science},
  volume       = {2045},
  pages        = {453--474},
  publisher    = {Springer},
  year         = {2001}
}

@inproceedings{conf/crypto/BellareR93,
  author       = {Mihir Bellare and
                  Phillip Rogaway},
  editor       = {Douglas R. Stinson},
  title        = {Entity Authentication and Key Distribution},
  booktitle    = {{CRYPTO}},
  series       = {Lecture Notes in Computer Science},
  volume       = {773},
  pages        = {232--249},
  year         = {1993}
}

@inproceedings{conf/crypto/Krawczyk01,
  author       = {Hugo Krawczyk},
  title        = {The Order of Encryption and Authentication for Protecting Communications
                  (or: How Secure Is SSL?)},
  booktitle    = {{CRYPTO}},
  series       = {Lecture Notes in Computer Science},
  volume       = {2139},
  pages        = {310--331},
  publisher    = {Springer},
  year         = {2001}
}

@misc{rfc9257,
    series =    {Request for Comments},
    number =    9257,
    howpublished =  {RFC 9257},
    publisher = {RFC Editor},
    author =    {R. Housley and J. Hoyland and M. Sethi and C. A. Wood},
    title =     {{Guidance for External Pre-Shared Key (PSK) Usage in TLS}},
    pagetotal = 13,
    year =      2022,
    month =     jul
}

@misc{rfc9846,
    series =    {Request for Comments},
    number =    9846,
    howpublished =  {RFC 9846},
    publisher = {RFC Editor},
    author =    {E. Rescorla},
    title =     {{The Transport Layer Security (TLS) Protocol Version 1.3}},
    pagetotal = 137,
    year =      2026,
    month =     jul
}

@inproceedings{conf/asiacrypt/Namprempre02,
  author       = {Chanathip Namprempre},
  title        = {Secure Channels Based on Authenticated Encryption Schemes: {A} Simple
                  Characterization},
  booktitle    = {{ASIACRYPT}},
  series       = {Lecture Notes in Computer Science},
  volume       = {2501},
  pages        = {515--532},
  year         = {2002}
}

\appendix

\section{Ethics Considerations}
This work does not involve human subjects, personal data, user studies, or data derived from individuals. Our experiments are conducted in a controlled local setting using QKD-derived experimental data and locally implemented post-processing components. We do not perform
network scanning, interact with third-party production systems, exploit deployed services, or evaluate attacks against real users or organizations.

The paper proposes and evaluates a key-establishment architecture and does not report any vulnerability in an existing system. Accordingly, there is no responsible-disclosure process required for this work. The experimental results are intended to assess the feasibility and performance of the proposed design under the stated assumptions, rather than to certify a production deployment.

\section{Related Work}
\label{sec:related}

\subsection{QKD Networks}
QKD networks~\cite{peev2009secoqc,stucki2011long,sasaki2011field} rely on trusted repeaters, metropolitan backbones, and secret key
management to extend QKD beyond single‑link distances.
Satellite-assisted QKD~\cite{yin2017satellite,liao2018satellite} further extends geographic reach with space infrastructures. 
The standardization
of QKD networks has become an active research topic, driven by the
challenges of coordinating heterogeneous deployments and interfaces~\cite{Stanley_2022}.

Despite the progress,  existing QKD networks remain device-centric, where each participating node has to host specialized QKD hardware, maintain quantum channels, and execute post-processing. The architectures also require the intermediate nodes, \ie, relays or repeaters, to be trusted when routing the secret keys. 
These systems improve QKD coverage, but typically feature limited elasticity under multi-user demand. 
  
In contrast, \RAMP is a \emph{service-centric} framework, which only requires the service providers to possess QKD hardware. 
It decouples the key provision from its consumption, enabling more flexible consumption of QKD-derived material.

\subsection{Offloading and QKD-as-a-Service}
Offloading processing of raw-key materials has recently been studied~\cite{journals/entropy/LorunserKPS23,journals/cee/RamalingamRKL24}.
They consider partial delegation to a third party, rather than \emph{systematically offloading} it to users. 

A branch of QKD as a service (QaaS)~\cite{cao2019multi,cao2019sdqaas} studies how to arrange resources over existing optical networks via network orchestration or physical design to match the multi-user requirement. 
Cao~\etal~\cite{cao2019multi} consider multi-tenant secret-key assignment within QKD networks, where software-defined networking (SDN) and quantum key pools are used to allocate finite secret-key rates among tenant requests. 
The follow-up SDQaaS~\cite{cao2019sdqaas} further develops SDN-based QaaS functions for service creation, modification, routing, and assignment of secret-key rates over an existing QKD infrastructure.
These works improve the operational efficiency of QKD networks through resource management and service provisioning. They do not address the trust problem that arises when service providers may abuse the keys that were assigned to users.

Kozlovics~\etal~\cite{conf/ispec/KozlovicsPLV23} propose a service framework that enables users to obtain QKD keys from two key distribution centers (similar to the service providers in \RAMP) over classical networks. 
Their design deploys a butterfly-style delivery protocol in which the two centers send different halves of a QKD-generated key, ensuring that no single classical channel carries the complete key. 
Consequently, a user must communicate with both centers to recover the QKD keys. This mechanism strengthens key delivery over classical channels, but it does not hide the key from the key distribution centers.

Different from the prior works, \RAMP makes access to QKD raw-key materials as a service and lets users locally mask and process them into final keys. 
\RAMP resolves the trust challenge by incorporating quantum-safe cryptography and formalizing security against the service infrastructure.

\subsection{Hybrid PQC–QKD Key Establishment and Key Combination}
Hybrid constructions combine independently protected key sources so
that the resulting session key can remain secure when one source
of key fails~\cite{conf/pqcrypto/DowlingHP20,
conf/pqcrypto/BrucknerRS23,journals/jsac/GarciaASORM25}.
When deployed for QKD-as-a-service,  a straightforward adaption is for service providers to complete QKD post-processing, deliver a final QKD key to the users, and let the users combine it with an independently established pre- or post-quantum secret.
Such a design could provide analogous source-failure confidentiality as \RAMP while requiring less QKD processing at the users. 

To our knowledge, prior hybrid and QKD-service constructions do not address this pre-reconciliation service boundary or the associated problem of protecting provider-known noisy QKD inputs while preserving user-side reconciliation.
In \RAMP, providers expose pre-reconciliation QKD-derived material, and the users execute the classical QKD security pipeline locally. The problem is addressed through our reconciliation-preserving masking and formal
analysis of the resulting service boundary.

\section{More on Cryptographic Building Blocks}
\label{appx:pqc}

\subsection{Digital Signature Scheme (SIG)}
\label{appx:sig}

\smallskip\noindent 
\textit{Correctness.}
A digital signature scheme \SIG is correct if, for any security parameter $\secparam$, any $(pk,sk)\in\keygen(\secparam)$, and any message $m\in\mathcal{M}$ with signature $\sigma\in\sign(sk,m)$, 
it holds that $\verify(pk,m,\sigma)=1$ with overwhelming probability.

\smallskip\noindent 
\textit{Security.}
The standard security notion for signature schemes is \emph{existential unforgeability under chosen-message attacks} (EUF-CMA). 
It requires that even if an adversary is allowed to obtain signatures on arbitrary messages of its choice, it cannot produce a valid signature on any message that it did not query.

\begin{definition}[EUF-CMA Security]
\label{def:EUF-CMA}
\SIG is existentially unforgeable under chosen-message attacks 
if, for any probabilistic or quantum polynomial-time adversary $\mathcal{A}$, the following probability is negligible in $\lambda$:

\[
\Pr\Big[
\begin{array}{c}
(pk,sk)\gets\keygen(\secparam),\; \\
(m^*,\sigma^*) \gets \mathcal{A}^{\sign(sk,\cdot)}(pk) : \\
\verify(pk,m^*,\sigma^*)=1 \;\wedge\; m^* \notin \mathcal{Q}
\end{array}
\Big].
\]
where $\mathcal{Q}$ is the set of messages queried by $\mathcal{A}$ to its signing oracle. 
\end{definition}

\smallskip\noindent 
\textit{Quantum Resistance.}
In response to the quantum threat, 
NIST has approved ML-DSA~\cite{ml-dsa}, derived from CRYSTALS-Dilithium, as a post-quantum digital-signature standard.
ML-DSA relies on the computational difficulty of solving the module learning with errors problem~\cite{LangloisS15}, which is believed to be difficult for adversaries with quantum capabilities.

\subsection{Authenticated Key Exchange (AKE)}
\label{appx:ake}

\smallskip\noindent 
\textit{Security.}
In general, $\AKE$ is secure if it is infeasible for the adversary to distinguish a key generated by the protocol from an independent random value sampled from the key space $\mathcal{K}$.
\begin{definition}[Session-Key Security]
\label{def:sk}
Consider an adversary $\mathcal{A}$ that controls the communication
network and may interact concurrently with protocol sessions, subject to the exposure rules.
For a completed and unexposed target session $(\usr,\usr',\mathsf{sid})$ between two parties $\usr$ and $\usr'$, let $k$ denote its session key.
We define the following advantage $\mathsf{Adv}_{\AKE,\adv}^{\mathsf{SK}}(\secparam)$
\[
\left|
\Pr\Big[
\begin{array}{c}
b \xleftarrow{\$} \{0,1\},\\
k_0 \gets \AKE(U,U',\mathsf{sid}), \
k_1 \sample \mathcal{K}\\
b' \gets \mathcal{A}(k_b):
b'=b
\end{array}
\Big] - \frac12.
\right|
\]

We say that $\AKE$ provides session-key security if
(i) matching sessions completed by uncorrupted $\usr$ and $\usr'$ output the same
session key, and
(ii) $\mathsf{Adv}_{\AKE,\adv}^{\mathsf{SK}}(\secparam)$ is negligible in $\lambda$.
\end{definition}

\smallskip\noindent 
\textit{Quantum Resistance.}
For post-quantum \AKE, the established session key of an unexposed accepted session must remain computationally indistinguishable from an independently sampled key even when the adversary has quantum computational capability.
We instantiate post-quantum \AKE in our construction with mutually authenticated TLS 1.3 that employs NIST-standardized post-quantum primitives for key establishment~\cite{ml-kem} and peer authentication~\cite{ml-dsa}.

\subsection{Secure Channel}
\label{app:scpc} 

\smallskip\noindent 
\textit{Security.}
We adopt the standard secure-channel requirements of confidentiality
and network authentication/integrity~\cite{conf/crypto/Krawczyk01,conf/asiacrypt/Namprempre02}, where the provisioned symmetric credential replaces the online establishment of the channel trust root.

\begin{definition}[Secure Channel Security]
\label{def:sc-security}
Let $\mathsf{SC}$ be a stateful two-party channel protocol rooted in a
provisioned credential. We say that $\mathsf{SC}$ is secure if,
whenever the target credential remains unexposed, the channel provides:
\begin{enumerate}
    \item confidentiality of protected messages against an active
    network adversary; and
    \item channel authentication and integrity, such that an honest
    receiver does not accept an unauthorized modification, injection,
    or replay under the target channel context,
\end{enumerate}
according to the standard secure-channel notions for symmetric
authenticated communication
~\cite{conf/crypto/Krawczyk01,conf/asiacrypt/Namprempre02}.

We write
$\mathsf{Adv}^{\mathsf{SC\mbox{-}SEC}}_{\mathsf{SC},\adv}(1^\lambda)$
as shorthand for the aggregate advantage of $\adv$ against these confidentiality and integrity notions. $\mathsf{SC}$ is secure
if
$
\mathsf{Adv}^{\mathsf{SC\mbox{-}SEC}}_{\mathsf{SC},\adv}(1^\lambda)
$
is negligible in $\lambda$ for every probabilistic or quantum
polynomial-time adversary $\adv$.
\end{definition}

In our security analysis of \RAMP,
$\mathsf{Adv}^{\mathsf{SC\mbox{-} SEC}}_{\mathsf{SC},\adv}(1^\lambda)$
upper-bounds the combined distinguishing and integrity-failure loss
incurred by one target secure-channel execution.
The quantity is negligible when $\mathsf{SC}$ provides secure channel
security.

\smallskip\noindent 
\textit{Quantum Resistance.}
The trust root of SC could be a high-entropy symmetric credential established during the prior provisioning phase. 
Its resistance to quantum adversaries instead depends on the security parameters and symmetric primitives used by the instantiated channel.

A concrete instantiation is TLS~1.3 with an externally provisioned
PSK in PSK-only mode~\cite{rfc9846,rfc9257}. TLS~1.3 explicitly
supports externally established PSKs and derives fresh
connection-specific traffic keys from the shared PSK. 
Symmetric primitives with enough key sizes remain appropriate candidates against quantum-capable adversaries~\cite{chennistir}, which are also discussed in~\cref{appx:ssc}.

\subsection{Synchronous Stream Cipher (SSC)}
\label{appx:ssc}
\smallskip\noindent 
\textit{Correctness.}
Following~\cref{def:SSC}, 
\SSC is correct if, for any $\secparam$, any $k \in \keygen(\secparam)$, any output length $n$, any keystream $z\in \keystream(k,n)$, and any $n$-bit message $m$,
it holds that 
$\dec(\enc(m,z), z) = m$
with overwhelming probability.

\smallskip\noindent 
\textit{Security.}
The standard security notion for SSC is \emph{indistinguishability from random bits against chosen-plaintext attack} (IND-CPA). 
Intuitively, it requires that for any efficient adversary, a ciphertext is computationally indistinguishable from a truly random sequence of the same length, even if the adversary can choose the plaintext to be challenged with adaptively.
\begin{definition}[IND-CPA Security of \SSC]
\label{def:SSC-Sec}
An $\SSC=(\keygen,\keystream,\enc,\dec)$
is indistinguishable under chosen-plaintext attacks if, for any
probabilistic or quantum polynomial-time adversary~$\adv$, the following advantage $\mathsf{Adv}_{\SSC,\adv}^{\indcpa}(\secparam)$ is negligible in~$\lambda$:

\[
\left|
\Pr\Big[
\begin{array}{c}
k \gets \keygen(\secparam), \\
(m,n) \gets \mathcal{A}(),\;
b \xleftarrow{\$} \{0,1\},\\
z \gets \keystream(k,n),\;
c_0 \gets \enc(m,z),\\
c_1 \sample \{0,1\}^{|c_0|},\\
b' \gets \mathcal{A}(c_b):
b'=b
\end{array}
\Big] - \frac12.
\right|
\]
\end{definition}

\smallskip\noindent 
\textit{Quantum Resistance.}
While public-key primitives like RSA or ECC are broken by Shor’s quantum algorithm~\cite{focs/Shor94}, symmetric-key primitives (and stream ciphers built upon them) offer better resistance. 
The foremost quantum threat is 
Grover’s algorithm~\cite{stoc/Grover96}, which gives a square-root speed-up for brute-force key search with quantum computers. That is, an \(n\)-bit key can in theory be searched in approximate \(O(2^{n/2})\) quantum steps using a quantum computer. 
This also implies that doubling key size can restore the original security level. 

The report by NIST~\cite{chennistir} remarks that symmetric algorithms remain viable in a quantum future provided key lengths are increased appropriately.   
More recent analyses and quantum resource estimates~\cite{ETSI} refine this understanding by considering circuit depth, qubit overhead, and fault-tolerance constraints.
The consensus remains that the stream ciphers retain resilience against realistic quantum attacks, assuming proper implementations and avoidance of structural weaknesses.

We instantiate \SSC with AES in counter mode (AES-CTR). 
It is deemed quantum-safe with an appropriate key length~\cite{conf/pqcrypto/AnandTTU16}.

\section{Security Arguments of \RAMP}

\subsection{Proof of Provider-Compromise Final-Key Confidentiality (\cref{thm:pc})}
\label{appx-pc-proof}

We first show the pseudorandomness of the keystream generated in \SSC by \cref{lem:ssc-prg}.
\begin{lemma}[Pseudorandom Keystream]
\label{lem:ssc-prg}
Let $\SSC$ be an XOR-based synchronous stream cipher. If $\SSC$ is
\indcpa\ secure as in Definition~\ref{def:SSC-Sec}, then for any probabilistic or quantum polynomial-time adversary $\adv$ and any output length $n$ decided by $\adv$, the following advantage is negligible in $\lambda$:
\[
\left|
\Pr\Big[
\begin{array}{c}
k \gets \keygen(\secparam),\quad
b \xleftarrow{\$} \{0,1\},\\
z_0=\keystream(k,n),\quad z_1\sample\{0,1\}^n,\\
b' \gets \mathcal{A}(z_b):
b'=b
\end{array}
\Big] - \frac12.
\right|
\]
\end{lemma}

\cref{lem:ssc-prg} can be proved as follows.
Assume there exists an adversary $\adv$ that could distinguish
$\keystream(k,n)$ from a uniform string with non-negligible advantage.
We build an adversary $\mathcal{B}$ against the \indcpa\ security of $\SSC$.

$\mathcal{B}$ submits the plaintext $m=0^n$ and output length $n$ as the challenge to the challenger in \SSC. It receives $c$ that is either
\begin{align*}
\enc(0^{n},\keystream(k,n))&=0^{n}\oplus\keystream(k,n)\\
&=\keystream(k,n)
\end{align*}
or a uniform string, depending on the challenge bit. $\mathcal{B}$
forwards $c$ to $\adv$ and outputs whatever $\adv$ outputs.
Thus, any non-negligible distinguishing advantage of $\adv$ gives the same advantage against \indcpa\ security of $\SSC$. It thus contradicts the fact that $\SSC$ is
\indcpa\ secure.

\paragraph{Proof of~\cref{thm:pc}}
We prove it by a sequence of hybrid games.
Let $\mathcal A$ be an admissible adversary in
$\mathsf{PC}_{\Pi, \adv}(1^\lambda)$.
For each hybrid $\mathsf H_i$, let
$p_i=\Pr[\mathcal A \text{ outputs } b'=b
\text{ in } \mathsf H_i
].
$

\noindent
\textbf{$\mathsf H_0$: }
$\mathsf H_0$ is exactly the real experiment of \cref{def:PC}.
Both providers may be corrupted and may expose their complete
target-session state. The honest users establish
$
k_{\mathsf{AKE}}
\gets
\mathsf{AKE}(U,U',\mathsf{sid}),
$
derive
$
z
\gets
\mathsf{SSC.Keystream}(k_{\mathsf{AKE}},n),
$
and execute masking and post-processing as in \cref{sec:construction}.

By definition, we have 
$$
\mathsf{Adv}^{\mathsf{PC}}_{\Pi,\adv}(1^\lambda)
=
\left|
p_0-\frac12
\right|.
$$

\noindent
\textbf{$\mathsf H_1$: } 
$\mathsf H_1$ is identical to $\mathsf H_0$, except that the common
key used by the target session for masking is replaced by an
independently sampled key
$
k^\star
\sample
\mathcal K,
$
where $\mathcal K$ is the \AKE/\SSC key space. The users derive
$
z
\gets
\mathsf{SSC.Keystream}(k^\star,n)
$
and execute the remaining parts of \RAMP unchanged.

Condition $\mathsf M$ guarantees that the target \AKE session is
unexposed. Therefore, if $\mathcal A$ distinguishes
$\mathsf H_0$ from $\mathsf H_1$, we obtain an adversary
$\mathcal B_1$ against the \AKE session-key security.

Specifically, $\mathcal B_1$ embeds its \AKE challenge key as the
masking key of the target \RAMP session and simulates the remaining
\RAMP execution for $\adv$. If the \AKE challenger supplies the
real session key, the simulation is distributed as $\mathsf H_0$;
if it supplies an independent key from $\mathcal K$, the simulation
is distributed as $\mathsf H_1$.

Under the advantage convention of~\cref{def:sk},

$$
|p_0-p_1|
\le
2
\operatorname{Adv}^{\mathsf{SK}}_{\mathsf{AKE},\mathcal B_1}
(1^\lambda).
$$

\noindent
\textbf{$\mathsf H_2$: }
$\mathsf H_2$ is identical to $\mathsf H_1$, except that the \SSC
keystream is replaced by
$
z^\star
\sample
\{0,1\}^{n}.
$

All subsequent masking and post-processing operations use
$z^\star$.
Since $k^\star$ is independently distributed as an \SSC key, \cref{lem:ssc-prg}
implies that replacing
$
\mathsf{SSC.Keystream}(k^\star,n)
$
by a uniform $n$-bit string is computationally
indistinguishable. Hence there exists an \indcpa\ adversary
$\mathcal B_2$ such that

$$
|p_1-p_2|
\le
2
\mathsf{Adv}^{\indcpa}
_{\SSC,\mathcal B_2}
(1^\lambda).
$$

It remains to analyze the information-theoretic part of
$\mathsf H_2$.
We consider the two QKD modes separately.
\begin{itemize}
    \item \textbf{DV-QKD.} 
$\usr'$ holds the binary
reference string $x'\in\{0,1\}^{n}$ before starting reconciliation.
After masking in $\mathsf H_2$, $\usr'$ has 
$
\rkm'= x'\oplus z^\star.
$

The provider-compromise experiment permits $\adv$ to know
$x'$ completely. This does not reduce the entropy introduced by the
mask. Since $z^\star$ is uniform and independent of all information
available when $x'$ is fixed,
$\rkm'$ is itself uniformly distributed over $\{0,1\}^{n}$, even conditioned
on the complete provider-side view.
    \item \textbf{CV-QKD.}
    In the reverse-reconciled CV-QKD, as described in~\cref{sec:mask},
    $U'$ holds the binary reference string $x'\in\{0,1\}^{n}$, while $U$ holds its corresponding LLR vector $x=(L_1,\ldots,L_n)$. 
    \RAMP transforms them as
    $$
    rk'_{\mathsf{mask}}
    =
    x'\oplus z^\star, \quad
    \widetilde L_i
    =
    (-1)^{z_i^\star}L_i.
    $$
    
As in the DV case, the binary reference
$x'\oplus z^\star$ is uniformly distributed even if the corrupted
providers know the original $x'$ and $L$ completely.

The transformed LLR vector $\widetilde L$ is auxiliary soft
information used locally by the decoder; it is not released to the
service providers or included in the public transcript. Its role is
to preserve the likelihood relation associated with the masked
reference bits: flipping $x'_i$ is accompanied by exchanging the
corresponding binary likelihood labels, represented by negating
$L_i$. Thus, the CV transformation does not supply additional
entropy; rather, it ensures that reconciliation remains consistent
with the uniformly masked binary reference string.
\end{itemize}

Therefore, for both QKD modes, the value from which the
reconciled key is derived contains an $n$-bit uniformly masked binary
reference string.

More generally, let $E$ denote the adversary's complete state
immediately after masking and before the subsequent key-dependent
public communication. The state $E$ may contain $x'$, the original
raw-key materials, metadata, acquisition credentials, the AKE
transcript, and all earlier public messages. In $\mathsf H_2$,
$
H_{\min}
\left(
\rkm'
\mid E
\right)
=
n.
$

This equality does not arise from the QKD source. It follows solely
from the independent uniform mask $z^\star$.

Let $C$ denote all key-dependent public information subsequently
released by reconciliation and correctness verification. By the
output-length required in \RAMP (\cref{cons:processing}), this public leakage is
accounted for before privacy amplification. Equivalently, if
$\mathsf{leak}$ denotes the corresponding conditional-min-entropy
loss, then
$$
H_{\min}
\left(
\rkm'
\mid E,C
\right)
\ge
n-\mathsf{leak}.
$$

Conditioned on successful reconciliation, both honest users hold the
same masked reference value on which privacy amplification is applied.
By the assumptions in~\cref{thm:pc}, the universal-hash privacy-amplification
output length $\ell$ is selected after accounting for this
post-masking leakage. With the leftover-hash guarantee the resulting final key is $\varepsilon_{\mathsf{PA}}$-close to an independent uniformly random $\ell$-bit string.
Consequently, replacing the real final key in $\mathsf{H}_2$ with an
independent uniform string changes the adversary's view by at most
$\varepsilon_{\mathsf{PA}}$.

Combining the bounds by the triangle inequality yields
$$
\begin{aligned}
\mathsf{Adv}^{\mathsf{PC}}_{\Pi,\adv}(1^\lambda)
&=
\left|
p_0-\frac12
\right|
\\
&\le
|p_0-p_1|
+
|p_1-p_2|
+
\left|
p_2-\frac12
\right|
\\
&\le
2
\mathsf{Adv}^{\mathsf{SK}}_{\AKE,\mathcal B_1}(1^\lambda)+
2\mathsf{Adv}^{\indcpa}_{\SSC,\mathcal B_2}(1^\lambda)
+
\varepsilon_{\mathsf{PA}}
\end{aligned}
$$

The first two terms are negligible by the assumed post-quantum
security of \AKE and \SSC, while
$\varepsilon_{\mathsf{PA}}$ is negligible for the selected
privacy-amplification parameters. Hence
$\mathsf{Adv}^{\mathsf{PC}}_{\Pi,\adv}(1^\lambda)$
is negligible, proving provider-compromise final-key confidentiality.

Importantly, this argument does not rely on QKD-secrecy or acquisition-security assumptions. The adversary may know the complete unmasked
QKD-derived inputs. In the provider-compromise branch, the entropy
used for final-key extraction is introduced by the independent
user-side mask, while privacy amplification removes the bounded
information subsequently exposed through public post-processing.
\hfill$\square$

\subsection{Proof of Masking-Compromise Final-Key Confidentiality (\cref{thm:mc})}
\label{appx-mc-proof}

We prove \cref{thm:mc} in two steps. 
We first account for the confidentiality and integrity of the two
target acquisition channels using the secure-channel security. Specifically, we use
$\mathsf{Adv}^{\mathsf{SC\mbox{-}SEC}}_{\mathsf{SC},\adv}(1^\lambda)$
as defined in \cref{def:sc-security} to represent the security loss
of each target acquisition channel.

We then show that, once the masking state is revealed,
\RAMP masking constitutes only a publicly invertible relabeling of the
underlying QKD post-processing execution.

Recall that the masking-compromise experiment exposes the complete
cryptographic masking state, including $k_\AKE$ and $z$,
but does not compromise either user endpoint. Conditions $\mathsf Q$
and $\mathsf A$ remain in force. Thus, the QKD-derived materials are
not disclosed by the service providers and are delivered to the users
without adversarial disclosure or modification.

The proof does not need the secrecy or pseudorandomness of the
masking state. Instead, it uses the fact that a disclosed mask induces
a public relabeling of the reconciliation inputs.

\begin{lemma}[Public-Mask Invariance]
\label{lem:invar}
Conditioned on the public session context, the masking state, including the masking keystream $z$, is
generated independently of the secret contents of the QKD-derived
reconciliation inputs. 

For DV-QKD masking~(\cref{sec:mask}),
$
(x,x')
\longmapsto
(x\oplus z,x'\oplus z)
$
is an invertible relabeling that preserves the complete discrepancy
pattern.

For CV-QKD masking~(\cref{sec:mask}),
$
x'
\longmapsto
x'\oplus z
$,
together with
$
L_i
\longmapsto
(-1)^{z_i}L_i
$,
is an invertible relabeling of the corresponding binary
soft-decision reconciliation problem.

If the instantiated reconciliation procedure is equivariant under
these relabelings, then the masked and unmasked executions have the
same accept/abort behavior and the same reconciliation-leakage bound,
and their reconciled binary reference strings differ only by the
public XOR offset $z$.
\end{lemma}

\paragraph{Proof of~\cref{lem:invar}.}
We analyze the QKD modes separately.
\begin{itemize}
    \item \textbf{DV-QKD.} 
For any fixed $z\in\{0,1\}^{n}$, applying the DV masking operation
twice gives
$$
(x\oplus z)\oplus z=x, \quad
(x'\oplus z)\oplus z=x'.
$$
Hence the transformation is a bijection whose inverse is itself.
Moreover,
$
(x\oplus z)\oplus(x'\oplus z)
=
x\oplus x'.
$
Namely, every disagreement position between the two reconciliation
inputs is preserved exactly. In particular, the error pattern and
error rate seen by the reconciliation procedure are unchanged.

\item \textbf{CV-QKD.} 
In the reverse-reconciled construction, $U'$ holds the binary
reference string $x'$, while $U$ holds the corresponding LLR vector
$
x=(L_1,\ldots,L_n).
$
Applying the same transformation twice restores both values as
$$
(x'_i\oplus z_i)\oplus z_i=x'_i, \quad 
(-1)^{z_i}(-1)^{z_i}L_i=L_i.
$$

The reliability of every soft-decision position is unchanged.
When $z_i=1$, the binary label is exchanged and the corresponding
LLR sign is reversed; when $z_i=0$, neither is changed. 
The same binary likelihood problem thus remains under
a different labeling after masking.

For the linear syndrome reconciliation in CV-QKD, 
if $\mathbf H$ is the parity-check matrix, then
$$
\mathbf H(x'\oplus z)
=
\mathbf Hx'
\oplus
\mathbf Hz.
$$
Hence the syndrome of the masked reference differs from the original
syndrome only by the public offset $\mathbf Hz$. Under the assumed
reconciliation equivariance, decoding the sign-relabeled LLRs with
this relabeled syndrome outputs the correspondingly relabeled
reconciled binary string and preserves the decoding success or failure.
\end{itemize}

Therefore, for either QKD mode, knowing $z$ permits the masked
reconciliation execution to be mapped efficiently and reversibly to
the corresponding unmasked execution. The transformation changes
neither the reconciliation acceptance behavior nor its leakage bound.
\hfill$\square$

\smallskip\noindent 
\textit{Instantiation to the evaluated CV-QKD pipeline.}
The CV-QKD implementation in~\cref{sec:real_qkd_eval} uses a binary reference
string, LLR-based belief-propagation decoding, and linear LDPC
syndromes. Hence its masking transformation follows the CV
instance of~\cref{lem:invar}. The LLR sign change exchanges the two binary
likelihood labels without changing their magnitudes, and the
syndrome changes only by the public offset.
Correctness-verification messages and the resulting acceptance bit
are treated as part of the public reconciliation transcript and are
included in the leakage accounting. Privacy amplification is applied
to the reconciled binary string.

\RAMP does not re-derive the finite-size security analysis of the
underlying CV-QKD protocol.
Provided the instantiated QKD protocol supplies a valid finite-size secrecy bound, \cref{lem:invar} shows that the relabeling in \RAMP preserves the reconciliation structure and leakage quantities to which that bound applies.

\paragraph{Proof of~\cref{thm:mc}.}
We prove it by a sequence of hybrid games.
Let $\adv$ be an admissible adversary in $\mathsf{MC}_{\Pi,\adv}(1^\lambda)$.
For each hybrid $\mathsf H_i$, let
$p_i=\Pr[\mathcal A \text{ outputs } b'=b
\text{ in } \mathsf H_i
].
$

\noindent
\textbf{$\mathsf H_0$: }
$\mathsf H_0$ is exactly the masking-compromise experiment of
\cref{def:MC}. The users acquire the target raw-key materials and
metadata through the real secure channels instantiated from
their independently provisioned credentials.

By definition, we have 
$$
\mathsf{Adv}^{\mathsf{MC}}_{\Pi,\adv}(1^\lambda)
=
\left|
p_0-\frac12
\right|.
$$

\noindent
\textbf{$\mathsf H_1$: }
$\mathsf H_1$ is identical to $\mathsf H_0$, except that the
acquisition channels between $\usr$ and $\sp$ is replaced according to the experiment of secure channel security.
Specifically, for the network-visible protection of the target acquisition payload,
the protected raw-key material and metadata are replaced by
equal-length dummy values. The honest endpoint nevertheless receives
the original provider-supplied payload internally. Executions in
which the adversary causes $\usr$ to accept an unauthorized modification, injection, or replay on this channel are treated as a secure-channel integrity failure.

If $\adv$ distinguishes $\mathsf H_0$ from $\mathsf H_1$, or
causes the accepted acquisition payload to differ from the
provider-supplied payload, we obtain an adversary $\mathcal B_1$
against the secure-channel security (\cref{def:sc-security}). Therefore,

$$
|p_0-p_1|
\le
\mathsf{Adv}^{\mathsf{SC\mbox{-}SEC}}_{\mathsf{SC},\mathcal B_1}(1^\lambda).
$$

\noindent
\textbf{$\mathsf H_2$: }
Similar to $\mathsf H_1$, $\mathsf H_2$ further applies the same transformation to the acquisition path between $P'$ and $U'$.

Thus, there exists an adversary $\mathcal B_2$ such that
$$
|p_1-p_2|
\le
\mathsf{Adv}^{\mathsf{SC\mbox{-}SEC}}_{\mathsf{SC},\mathcal B_2}(1^\lambda).
$$

We now analyze $\mathsf H_2$. 
Condition $\mathsf Q$ guarantees that the target QKD execution
satisfies the secrecy assumptions of the instantiated QKD protocol
and that the provider-side state containing the target raw-key
materials is not exposed to $\adv$. 

Consequently, immediately before masking, the secret-bearing QKD
state held by the users has the same secrecy guarantee as in the
corresponding underlying QKD execution. 
Let $\mathsf{Aux}_{\mathsf{QKD}}$ denote the side information covered
by the composable QKD secrecy definition, including the adversarial
quantum side information and the public transcript of the underlying
QKD classical processing.

By the two secure-channel hybrids above, the acquisition traffic in
$\mathsf H_2$ contributes no information about the protected
QKD-derived contents beyond permitted public channel information, and
the users' accepted acquisition payloads equal the corresponding
provider-supplied payloads except for the secure-channel failure
probabilities already accounted for.

The masking-compromise experiment then reveals both
$k_{\AKE}$
and
$z$ to $\adv$, thus no secrecy of \AKE or \SSC is used.

By construction, the reconciliation inputs are fixed before the users
execute \AKE and derive $z$. Hence revealing the masking state does not
reveal an additional function of the user-side secret QKD-derived inputs.
For every resulting public value of $z$, \cref{lem:invar} shows that the \RAMP
masking operation is an invertible relabeling of the reconciliation
problem.

It remains to verify that privacy amplification preserves the same
security guarantee after this relabeling.
Let
$
\tau_z(v)=v\oplus z
$
denote the relabeling of the reconciled binary reference string.
\cref{cons:processing} applies privacy amplification using a universal hash
family. Let this family be
$
\mathcal H
=
\{h_s\}_s.
$
Executing privacy amplification on the masked reconciled string is
equivalent to applying
$
h_s\circ\tau_z
$
to the corresponding unmasked string.

For every fixed $z$, the composed family
$
\mathcal H_z
=
\{h_s\circ\tau_z\}_s
$
remains universal. 
Specifically, for any distinct
$v,v'\in\{0,1\}^{n}$,
$
\tau_z(v)
\neq
\tau_z(v'),
$
and therefore
$$
\Pr_s[
h_s(\tau_z(v))
=
h_s(\tau_z(v'))
]
\le
2^{-\ell}.
$$
Thus, the public XOR relabeling does not weaken the
privacy amplification. Together with~\cref{lem:invar}, the masked
execution has the same reconciliation leakage and the same
privacy-amplification security conditions as the corresponding
unmasked QKD execution.

For the LLR transformation in CV-QKD, the LLRs are auxiliary soft information used only to perform reconciliation, while the privacy amplification is applied to the resulting reconciled binary reference string. The sign transformation ensures that this binary string is exactly the relabeled version required by $\tau_z$.

Let $E_{\mathsf{qkd}}$ denote the adversarial side-information
system covered by the composable QKD secrecy definition, including
the quantum side information by $\adv$ and the public transcript
of the underlying QKD classical processing. Let
$E_{\RAMP}$ additionally include the public \RAMP session
context, the exposed masking state, and the post-masking public transcript.
By public-mask invariance and the preservation of universality under
the public relabeling $\tau_z$, the \RAMP post-processing execution
has the same secrecy guarantee as the corresponding underlying QKD
execution. 

We denote the subnormalized accepted states as
$${\rho}^{\mathsf{acc}}_{KE} = p_{\mathsf{acc}}\
\bar{\rho}^{\mathsf{acc}}_{KE},\quad {\rho}^{\mathsf{acc}}_{E} = p_{\mathsf{acc}}\
\bar{\rho}^{\mathsf{acc}}_{E} $$

Hence, for the subnormalized accepted branch,
$$
D\!\left(
\rho^{\mathsf{acc}}_{K E_{\mathsf{RAMP}}},
\tau_{K}
 \otimes
\rho^{\mathsf{acc}}_{E_{\mathsf{RAMP}}}
\right)
\le
\varepsilon_{\mathsf{qkd}}.
$$

Thus the \RAMP key is
$\varepsilon_{\mathsf{qkd}}$-close to a uniformly random key independent of the adversary's complete side information.

Combining the prior hybrid with the QKD secrecy bound, we have
$$
\begin{aligned}
\mathsf{Adv}^{\mathsf{MC}}_{\Pi,\mathcal A}(1^\lambda)
&=
\left|p_0-\frac12\right|
\\
&\le
|p_0-p_1|
+
|p_1-p_2|
+
\left|p_2-\frac12\right|
\\
&\le
\mathsf{Adv}^{\mathsf{SC\mbox{-}SEC}}_{\mathsf{SC},\mathcal B_1}(1^\lambda)
+
\mathsf{Adv}^{\mathsf{SC\mbox{-}SEC}}_{\mathsf{SC},\mathcal B_2}(1^\lambda)
+
\varepsilon_{\mathsf{qkd}}.
\end{aligned}
$$
Therefore, \RAMP provides masking-compromise final-key confidentiality
when $\mathsf Q\land\mathsf A$ holds.
\hfill$\square$

\subsection{Dual-Domain Compromise Resilience} \label{appx:dual-domain} 
The preceding theorems directly realize the confidentiality boundary defined in~\cref{subsec:threat-model}. 
\begin{corollary}[Dual-Domain Confidentiality] 
\RAMP realizes final-key confidentiality, provided that $ \mathbf M \lor (\mathbf Q\land\mathbf A) $ holds. 

If $\mathbf M$ holds, \cref{thm:pc} applies even when the complete provider-side QKD and acquisition state is exposed. If $\mathbf M$ does not hold but both $\mathbf Q$ and $\mathbf A$ hold, \cref{thm:mc} applies despite complete exposure of the masking state. 
\end{corollary} 

The two cases rest on different security foundations. Under provider compromise, \RAMP provides computational confidentiality through \AKE, \SSC, and privacy amplification. 
Under masking compromise, \RAMP retains the secrecy guarantee supplied by the underlying QKD execution, provided the QKD-derived materials remain protected at the service boundary and during acquisition. 
\RAMP makes no final-key confidentiality claim when $ \neg\mathsf M \land (\neg\mathsf Q\lor\neg\mathsf A). $ This is the explicit confidentiality boundary of \RAMP.

\section{More on QKD Experiment}

\subsection{QKD Testbed}

The qOptica~100 platform implements a Gaussian-modulated coherent-state CV-QKD protocol with heterodyne detection and reverse reconciliation, consistent with practical long-distance CV-QKD systems~\cite{jouguet2011long}. The transmitter and receiver are connected through a deployed metropolitan fibre loop of $13.7$\,km, so the acquired raw-key material reflects transmission over real installed field fibre rather than a laboratory spool, back-to-back, or simulated channel. 
The one-way propagation delay over the $13.7$\,km loop ($\approx 68$\,$\mu$s) and the syndrome-serialization latency are negligible relative to the user-side processing times reported.

CV-QKD needs to specify classical side information, \ie, metadata, required for post-processing.
This includes:
\begin{itemize}
\item Shot-noise calibration that is used to normalize quadrature measurements.
\item Detector parameters, \eg, detection efficiency $\eta_B$ and trusted electronic noise~$v_{\mathrm{el}}$.
\item Channel estimates, \eg, transmittance $T$ and excess noise $\xi$, obtained during parameter estimation.
\item Reconciliation configuration, \eg, LDPC code structure, code rate $R$, and block~size.
\item Protocol metadata, \eg, basis information (if applicable), block partitioning, and parameter estimation splits.
\end{itemize}
The providers supply calibration and protocol metadata, including shot-noise normalization, detector parameters, code configuration, and block-partition information. 
Security-critical channel quantities such as transmittance $T$ and excess noise $\xi$ are estimated by the users locally during the parameter estimation rather than trusted as provider-supplied values. 
The authentication of provider metadata guarantees its origin and transport integrity, which does not imply correctness under provider corruption.

The metadata is required for the construction of log-likelihood ratios (LLRs) and the correct operation of reconciliation and privacy amplification. In our implementation, quadrature samples are pre-normalized with respect to shot noise; the remaining parameters are still required to correctly model the effective channel and compute soft information.

\subsection{Post-Processing Details}
\label{appx:config}
Efficient error correction in CV-QKD requires transforming continuous Gaussian variables into a form amenable to binary error-correcting codes. We employ multidimensional reconciliation using 8-dimensional (octonion-based) mappings, following Leverrier \emph{et al.}~\cite{leverrier2008multidimensional}, in which blocks of heterodyne measurements are mapped via orthogonal transformations to an equivalent binary-input additive white Gaussian noise (BiAWGN) channel. This produces soft information in the form of LLRs, which serve as input to a syndrome-based LDPC decoder. Error correction uses large, sparse quasi-cyclic LDPC codes of block length $n_H = 2\times10^6$ bits and rate $R=1/10$, obtained from the qOptica QKD source files, with belief-propagation decoding on soft inputs. The effective channel model incorporates shot noise (normalised to unity), trusted detector electronic noise, and excess noise estimated during parameter estimation, which together determine the SNR and reconciliation performance.

The full pipeline explicitly realizes parameter estimation, multidimensional reconciliation, LDPC decoding, verification, and privacy amplification. We denote the user who processes data from the QKD transmitter (\resp receiver) by $\usr$ (\resp $\usr'$). In \RAMP, masking is applied prior to reconciliation by flipping both the raw bits of $\usr'$ and the corresponding LLRs of $\usr$ using a shared keystream, ensuring that decoding performance is unaffected while hiding the masked keys from the providers. Correctness is enforced via Toeplitz-hash verification, and privacy amplification is performed after aggregating multiple successfully decoded blocks to mitigate finite-size effects.

For reproducibility, we make explicit three implementation choices that determine the efficiency of the user-side pipeline:
\begin{itemize}
\item \emph{Vectorised belief propagation.} The decoder operates on the Tanner-graph edge list using vectorised check- and variable-node updates, rather than scalar per-node loops. This removes the dominant cost of the reference decoder while reproducing identical decoding behaviour (and hence identical FER).
\item \emph{Both-quadrature parameter estimation.} Channel parameters are estimated jointly over the amplitude and phase quadratures across the full parameter-estimation sample budget, tightening the excess-noise estimate and thereby the Holevo bound $\chi_{E|B}$.
\item \emph{FFT-based privacy amplification.} The Toeplitz hash is evaluated by blocked FFT, giving an $O(n\log n)$ extractor that scales to aggregated inputs of order $10^9$ bits.
\end{itemize}
Near-capacity reconciliation in practice still requires careful handling of LLR scaling, numerical stability, and decoder convergence, which we address in the implementation.

\section{Findings in Experiment:\\
Asymmetry, Optimization, and Autonomy}
\label{appx:findings}

The breakdown of the processing time shown in \cref{tab:exp-breakdown} provides two insights on performance. First, due to the asymmetric user-side workload, a practical deployment of \RAMP can be strategically adjusted based on user-side capability to improve performance.
Second, since the dominant operation is LDPC reconciliation, optimized implementations and acceleration from commodity GPUs~\cite{milicevic2018quasi} are expected to enable speedups without changing the workflow in \RAMP.

Our evaluation also highlights a feature
of \RAMP, \ie, raw-key acquisition and final-key distillation are decoupled. 
Users can autonomously decide when and how much raw keys to process, provided they follow the post-processing procedure.

\subsection{Asymmetry in User-Side Processing Time}
\label{appx:asymmetry}

The processing cost of \RAMP is asymmetric between the two users in our CV-QKD instantiation. This asymmetry comes from the structure of syndrome-based reconciliation.

For masking in reverse reconciliation (\cref{sec:mask}), $\usr'$ holds the masked binary
string $\rkm'$, while $\usr$ holds the corresponding masked LLRs. 
$\usr'$ computes the syndrome as
$$
    s = {\mathbf H} \rkm' \bmod 2,
$$
where ${\mathbf H} $ is the LDPC parity-check matrix. 
This operation is essentially
a lightweight sparse matrix-vector multiplication. 

On the other hand, $\usr$ performs the heavyweight part of reconciliation.
It runs iterative LDPC decoding over soft inputs and verifies the decoded result.

This finding has direct implications for deployment. \RAMP does not require both users to have symmetric computational capability. 
When role assignment is configurable, 
a resource-constrained device can participate in the syndrome-generation side, while a more capable endpoint handles the decoding-intensive reconciliation. 
Such a role assignment is particularly relevant in
service-oriented settings, where users may operate 
laptops, mobile devices, or embedded IoT devices with heterogeneous
computational resources.

Therefore, the practicality of \RAMP should be understood not only from
the total processing time, but also from the per-user workload distribution. 
The computation is concentrated on one side, while masking and syndrome generation remain lightweight. 
It provides a useful deployment insight, \ie, \RAMP can support heterogeneous users by assigning the reconciliation-heavy role to the more capable endpoint.

\subsection{Optimization with Commodity Hardware}
\label{appx-GPU}
\cref{tab:exp-breakdown} identifies LDPC reconciliation as the bottleneck of our prototype, which is introduced by standard CV-QKD post-processing. 
Since our implementation does not utilize hardware acceleration, the result should be viewed as a conservative CPU baseline rather than an inherent limit of \RAMP.

Prior work on CV-QKD reconciliation~\cite{milicevic2018quasi} has shown that LDPC decoding is highly amenable to parallel acceleration on commodity GPUs (with a single NVIDIA GeForce GTX $1080$ GPU in their implementation), especially when the code structure is designed to improve memory access and message-passing efficiency. 
More recent work~\cite{PhysRevA.109.012604} further suggests GPU-accelerated rate-adaptive reconciliation for CV-QKD by optimizing the
storage of the parity-check matrix.

These observations indicate that the dominant cost observed in our evaluation is an optimizable implementation bottleneck of reconciliation, rather than a protocol-level overhead. 
More specifically, a cost-effective method toward overall performance improvement is to assign the decoding-heavy role identified in \cref{appx:asymmetry} to an endpoint with commodity computing resources like a GPU. 
The other endpoint could remain on the lightweight side.

Note that the optimization remains independent of specialized QKD hardware. With commodity GPUs for QKD post-processing available, end users could configure and consume the service with general-purpose computing platforms.

\subsection{On-Demand Key Consumption and User Autonomy}

In conventional QKD deployments, the parties that benefit from the final keys are also involved in the quantum-state generation and transmission. 
In \RAMP, service providers generate and maintain raw-key materials, while users decide when and how much of the
acquired materials to process into final keys.

This separation gives users greater autonomy. A user can acquire raw-key materials when the service is available and defer post-processing until local computation, energy, or application demand makes it appropriate. 
The amount of material processed in each session can also be adjusted according to the available resources and the desired key length, as long as it satisfies the conditions of QKD post-processing. For example, 
the users could process smaller raw-key batches in a resource-constrained device, while aggregating larger batches when strong computation resources are available.

Such deferred processing requires appropriate local protection of the acquired raw-key materials. Users should store raw-key blocks together with their metadata, key identifiers, and state information, and protect them using standard authenticated encryption or quantum-safe storage mechanisms where appropriate. The raw-key state must also remain consistent with the provider-side lifecycle, so that each block is used only once and eventually marked as consumed or expired.

From the perspective of service providers, \RAMP 
removes reconciliation and privacy amplification from the providers (\cf partial delegation in prior works~\cite{journals/entropy/LorunserKPS23,journals/cee/RamalingamRKL24}). 
Service providers can concentrate on operating QKD devices, maintaining raw-key inventories, and serving acquisition requests, rather than performing reconciliation and privacy amplification for every user session. 
In this way, service providers produce and manage QKD-derived materials as a backend resource, while users consume and process them on demand.

Therefore, \RAMP does not merely offload computation from providers to users. It upgrades the operational model of QKD, where raw-key generation, raw-key acquisition, and final-key distillation no longer need to occur as one tightly coupled procedure. 
This decoupling is central to making QKD-derived entropy available as a flexible service for heterogeneous users.

\section{Use of Generative AI}
The main body of this paper, 
including the conceptual design, technical construction, analyses, evaluation, and novel intellectual contributions, was produced by the authors.

Generative AI tools were used in supporting roles.
\emph{ChatGPT} was used to assist with language polishing. \emph{ChatGPT} and \emph{Claude} helped build the prototype and fix bugs during development. All suggested fixes from generative AI were manually reviewed, tested, and validated by the authors before being included. \emph{Grammarly} was used to check grammar and improve the presentation of the text.
The authors take full responsibility for the accuracy, correctness, and integrity of all material in the paper.

\end{document}